\documentclass[11pt]{article}

\usepackage[square,numbers,sort&compress]{natbib}

\usepackage{tablefootnote}
\usepackage[utf8]{inputenc} 
\usepackage[T1]{fontenc}    
\usepackage{hyperref}       
\usepackage{url}            
\usepackage{booktabs}       
\usepackage{amsfonts}       
\usepackage{nicefrac}       
\usepackage{microtype}      
\usepackage{tikz}
\usetikzlibrary{decorations.pathreplacing,calc}

\usepackage{verbatim}
\usepackage{amsthm}
\usepackage{amsfonts}
\usepackage[nonamelimits]{amsmath}
\usepackage{amssymb}
\usepackage{sansmath}
\usepackage{color}
\usepackage{xcolor}
\usepackage{bm}
\usepackage{bbm}
\usepackage{fullpage}
\usepackage[bottom]{footmisc}

\theoremstyle{plain}
\newtheorem{theorem}{Theorem}[section]

\newtheorem{lemma}[theorem]{Lemma}

\newtheorem{claim}[theorem]{Claim}
\newtheorem{problem}[theorem]{Problem}
\theoremstyle{definition}
\newtheorem{definition}[theorem]{Definition}

\usepackage{subfig}
\usepackage{graphicx}
\usepackage{listings}
\usepackage{enumitem}
\usepackage{tabularx}
\newtheoremstyle{named}{}{}{\itshape}{}{\bfseries}{.}{.5em}{\thmnote{#3}}
\theoremstyle{named}

\usepackage{tikz}
\usetikzlibrary{fit,calc}

\usepackage{tikz}
\usetikzlibrary{cd}

\DeclareSymbolFont{extraup}{U}{zavm}{m}{n}
\DeclareMathSymbol{\varheart}{\mathalpha}{extraup}{86}
\DeclareMathSymbol{\vardiamond}{\mathalpha}{extraup}{87}

\newcommand{\R}{\mathbb{R}}

\newcommand{\EE}{\mathbb{E}}

\newcommand{\mbx}{\mathbf{x}}
\newcommand{\mby}{\mathbf{y}}

\makeatletter
\newcommand*{\inlineequation}[2][]{%
  \begingroup
    \refstepcounter{equation}%
    \ifx\\#1\\%
    \else
      \label{#1}%
    \fi
    \relpenalty=10000 %
    \binoppenalty=10000 %
    \ensuremath{%
      #2%
    }%
    ~\@eqnnum
  \endgroup
}
\makeatother

\usepackage{threeparttable}

\usepackage{todonotes}

\usepackage{xspace}

\xspaceaddexceptions{[]\{\}}

\usepackage{hyperref}

\usepackage{thm-restate}

\usepackage[linesnumbered,ruled,vlined]{algorithm2e}
\SetKwComment{Comment}{$\triangleright$\ }{}

\newcommand{\eat}[1]{}

\makeatletter
\newcommand*{\rom}[1]{\expandafter\@slowromancap\romannumeral #1@}

\usepackage{multirow}

\usepackage{tablefootnote}
\usepackage{float}
\usepackage{makecell}

\allowdisplaybreaks

\title{On Optimal Approximations for $k$-Submodular Maximization via Multilinear Extension}
\author{
Lingxiao Huang \thanks{{\tt  huanglingxiao@nju.edu.cn}. Nanjing University.} 
\and
Baoxiang Wang\thanks{{\tt bxiangwang@cuhk.edu.cn}. The Chinese University of Hong Kong, Shenzhen.} 
\and
Huanjian Zhou\thanks{{\tt zhou@ms.k.u-tokyo.ac.jp}. Department of Complexity Science and Engineering, Graduate School of Frontier Sciences, The University of Tokyo.}
}

\date{}
\date{\vspace{-1ex}}

\begin{document}

\maketitle

\begin{abstract}
We investigate a more generalized form of submodular maximization, referred to as $k$-submodular maximization, with applications across social networks and machine learning domains. In this work, we propose the multilinear extension of $k$-submodular functions and unified Frank-Wolfe-type frameworks based on that. Our frameworks accomodate 1) monotone or non-monotone functions, and 2) various constraint types including matroid constraints, knapsack constraints, and their combinations. Notably, we attain an asymptotically optimal $1/2$-approximation for monotone $k$-submodular maximization problems with knapsack constraints, surpassing the previous $1/3$-approximation~\citep{ha4395206improved}. The foundation for our analysis stems from new insights into specific linear and monotone properties pertaining to the multilinear extension.
\end{abstract}

\tableofcontents
\newpage

\section{Introduction}
\label{sec:intro}

Consider the following problems in machine learning and operations research:
1) identifying influential individuals in a social network with $k$ topics to maximize the number of individuals influenced by at least one topic~\cite{qian2017constrained,zhang2019budget},
2) partitioning a set of features into $k + 1$ subsets such that one feature can be used in at most one regression target (or none of them) for $k$ regression targets on these features~\citep{DBLP:journals/jmlr/SinghGB12,zhou2019subset},
and 3) selecting a small set of sensors from $k$ types of sensors in an area to maximize the information obtained from the sensors~\citep{DBLP:conf/nips/OhsakaY15}.
%
%
These problems are often constrained, such as selecting sensors with different costs within a finite budget.

Solving these problems all involves maximizing a \emph{$k$-submodular} set function $f: \{0,\ldots, k\}^n\to \mathbb{R}_{\geq 0}$ subject to some constraints. 
Intuitively, the $k$-submodularity property captures the notion of diminishing returns in every orthant. 
%
For instance, consider identifying influential individuals in a social network. For a fixed topic, the newly selected influential individuals will contribute less to the overall user coverage if many influential individuals have already been selected, and more if only a few have been selected.
Similarly, adding additional features in regression problems and placing additional sensors in an area also share such diminishing returns property.

Formally, for an integer $k \geq 1$ and a finite nonempty set $[n]$, a function $f : \{0,\ldots,k\}^{n} \to \mathbb{R}_{\geq 0}$ is called {$k$-submodular} if for all $\mathbf{s}$ and $\mathbf{t}$ in $\{0,\ldots,k\}^{n}$, we have 
\[f(\mathbf{s})+f(\mathbf{t}) \geq f(\min_0(\mathbf{s},\mathbf{t})+f(\max_0(\mathbf{s},\mathbf{t})), \]
where for every $i\in [n]$,
\[\min_0(\mathbf{s},\mathbf{t})_i = \left\{\begin{array}{ll}    0, &  \mathbf{s}_i\mathbf{t}_i\neq 0,\mathbf{s}_i\neq \mathbf{t}_i,\\    \min(\mathbf{s}_i,\mathbf{t}_i), & \mbox{otherwise,}\end{array}\right.  \max_0(\mathbf{s},\mathbf{t})_i = \left\{\begin{array}{ll}
   0, &  \mathbf{s}_i\mathbf{t}_i\neq 0,\mathbf{s}_i\neq \mathbf{t}_i,\\
    \max(\mathbf{s}_i,\mathbf{t}_i), & \mbox{otherwise.}
\end{array}\right.\]


A special case of the $k$-submodular maximization problem is the submodular maximization problem, i.e., $k=1$. The techniques for submodular maximization problems can generally be classified into two main lines.
The first line is combinatorial and is mostly based on greedy rules and local search. This approach has been applied to both monotone and non-monotone submodular objective functions under various constraints~\citep{DBLP:journals/siamcomp/BuchbinderFNS15,DBLP:journals/siamcomp/FeigeMV11,filmus2014monotone,lee2010maximizing,DBLP:journals/mor/LeeSV10,nemhauser1978analysis}. In some cases, optimal algorithms have been obtained using this line of approaches~\citep{DBLP:journals/siamcomp/BuchbinderFNS15,DBLP:journals/orl/Sviridenko04}.
The second line is a two-staged framework based on the multilinear extension. 
This line of methods involves identifying a fractional solution for the relaxation of the problem and then rounding the fractional solution to obtain an integral one while incurring a bounded loss in the objective.
%
This line of approaches achieves better approximation ratios in most cases~\citep{DBLP:journals/corr/BuchbinderF16,DBLP:journals/siamcomp/CalinescuCPV11,DBLP:conf/focs/ChekuriVZ10,DBLP:journals/siamcomp/ChekuriVZ14,DBLP:conf/focs/FeldmanNS11,kulik2013approximations}.

Previous works in constrained $k$-submodular function maximization were based on combinatorial techniques, such as the greedy algorithm. 
%
%
%
However, compared with tight approximations of submodular maximization with various constraints, previous combinatorial approaches have not been able to achieve asymptotically optimal approximation results in most cases. In fact, the only tight results available are on the basic case of a single matroid constraint.
For example, even for the important case of monotone $k$-submodular maximization with single knapsack constraint, the current best combinatorial method only obtains $1/3$-approximation~\citep{ha4395206improved}, having a large gap with the previous lower bound $\frac{k+1}{2k}$~\citep{DBLP:conf/soda/IwataTY16}.
Also, combinatorial methods do not provide the flexibility to combine constraints of different types, especially for $O(1)$ knapsack constraints.

\begin{table}[t!]
\renewcommand\arraystretch{1.5}
\centering
\begin{tabular}{|c|c|c|c|}
\hline
\multicolumn{2}{|l|}{{Problem type of  $k$-submod. max.}} & 
\makecell[c]{Prior results}&\makecell[c]{Our results} \\
\hline
\multirow{2}{*}{$O(1)$ knapsacks} & \makecell[c]{Monotone} &  $\frac{1}{3}$~\citep{ha4395206improved}$^\clubsuit$    & $\mathbf{\frac{1}{2}-o(1)}^\spadesuit$\\
\cline{2-4}
  & \makecell[c]{Non-monotone}~($k\geq 2)$ & $\frac{1-e^{-4}}{4}$~\citep{yu2023maximizing}$^\clubsuit$    & $\mathbf{\frac{1}{3}-o(1)}$ \\
\hline
\multirow{2}{*}{single matroid}   &\makecell[c]{Monotone } & $\frac{1}{2}$~\citep{sakaue2017maximizing}$^\spadesuit$ &$\mathbf{\frac{1}{2}-o(1)}^\spadesuit $\\
\cline{2-4}
  & \makecell[c]{Non-monotone}~($k\geq 2)$   & $\frac{1-e^{-4}}{4}$~\citep{yu2023maximizing}     &  $\mathbf{\frac{1}{3}-o(1)}$\\
\hline
\multirow{2}{*}{\makecell[c]{$b$ matroids + \\  $O(1)$ knapsacks} }  & \makecell[c]{Monotone } & \makecell[c]{$\frac{1-e^{-(b+2)}}{b+2}$~\citep{yu2023maximizing}$^\clubsuit$ } & $\mathbf{\frac{0.3}{b}-o(1)}$ \\
\cline{2-4}
  & \makecell[c]{Non-monotone}~($k\geq 2)$   & $\frac{1-e^{-(b+3)}}{b+3}$~\citep{yu2023maximizing}$^\clubsuit$     & $\mathbf{\frac{0.2}{b}-o(1)}$ \\
\hline
\end{tabular}
\caption{Comparison with previous work for constrained $k$-submodular maximization. 
Symbol $^\clubsuit$ represents that the prior works \citep{ha4395206improved,yu2023maximizing} only consider a single knapsack instead of $O(1)$.
Symbol $^\spadesuit$ represents that the results are asymptotically tight.
}
\label{table:results}
\end{table}

\subsection{Our contributions}
In this paper, we propose unified frameworks for the problem of constrained $k$-submodular maximization (Problem \ref{prob}) via a novel multilinear extension (Definition~\ref{def:mul}).
We consider two classic types of constraints: matroid constraints (Definition~\ref{def:matroid}) and knapsack constraints (Definition~\ref{def:knapsack}).
Our results are summarized in Table \ref{table:results}.
We first present the results when $f$ is \emph{monotone}, i.e., if $f(\mathbf{s}) \leq f(\mathbf{t})$ holds for every pair of integral vectors $\mathbf{s},\mathbf{t}\in \{0,\ldots,k\}^n$ satisfying 1) $\mbox{supp}(\mathbf{s})\subseteq \mbox{supp}(\mathbf{t})$, where $\mbox{supp}(\mathbf{s}) := \{e\in [n]:\mathbf{s}_e\neq 0\}$ represents the support set, and 2) $\mathbf{s}_e = \mathbf{t}_e$ for all $e\in \mbox{supp}(s)$.

\begin{theorem}[\bf{Informal, see Theorem \ref{the:main}}]
\label{thm:main_informal}
For monotone constrained $k$-submodular maximization, there exists a (randomized) polynomial-time algorithm that returns 1) $1/2-o(1)$ approximation for $O(1)$ knapsacks or single matroid 2) $0.3/b-o(1)$ approximation for the intersection of $O(1)$ knapsacks and $b$ matroids.
\end{theorem}

For a single knapsack constraint, we improve the approximation ratio from $1/3$, as demonstrated in \citep{ha4395206improved}, to $1/2-o(1)$. 
Our approximation ratio matches the previous lower bound $\frac{k+1}{2k}$~\citep{DBLP:conf/soda/IwataTY16} and hence, is asymptotically tight.
Furthermore, we extend this result to the case of $O(1)$ knapsack constraints. 
For a single matroid constraint, we also obtain an asymptotically optimal approximation ratio $1/2-o(1)$, matching the previous work~\citep{sakaue2017maximizing}.
Moreover, our algorithm can handle the intersection of $O(1)$ knapsacks and $b$ matroids with an approximation ratio $\frac{0.3}{b} - o(1)$, which extends~\citep{yu2023maximizing}. 
We remark that factor $1/b$ is necessary due to a lower bound of $\Omega(\frac{\log b}{b})$ (Appendix~\ref{sec:hardness}).

We then present the theoretical results when $f$ is non-monotone.

\begin{theorem}[\bf{Informal, see Theorem \ref{the:nonmonotone}}]
\label{thm:nonmonotone_informal}
For non-monotone constrained $k$-submodular maximization where $k\geq 2$, there exists a (randomized) polynomial-time algorithm that returns 1) $1/3-o(1)$ approximation for $O(1)$ knapsacks or single matroid 2) $0.2/b-o(1)$ approximation for the intersection of $O(1)$ knapsacks and $b$ matroids.
\end{theorem}

We present improved approximation ratios for the non-monotone objective in the context of knapsack and matroid constraints. 
Specifically, for a single knapsack constraint or a single matroid constraint, we improve the approximation ratio from $(1-e^{-4})/4$, as demonstrated in \citep{ha4395206improved}, to $1/3-o(1)$. 
Furthermore, we extend this result to the case of $O(1)$ knapsack constraints. 
We also extend the result from the intersection of single knapsack and $b$ matroids case presented in \citep{yu2023maximizing} to the intersection of $O(1)$ knapsacks and $b$ matroids, with only a small loss in the approximation ratio.

\subsection{Technical overview}
\label{sec:overview}

We adopt a continuous approach for our algorithms (Algorithms~\ref{alg:main} and \ref{alg:nonmonotone}), utilizing the newly proposed multilinear extension (Definition~\ref{def:mul}). Our algorithms extend the previous ones \citep{DBLP:conf/aistats/BianMB017,DBLP:journals/siamcomp/CalinescuCPV11} for submodular maximization, for which we briefly review now.

\paragraph{Continuous methods for submodular maximization via multilinear extension.} 
Multilinear extension $F$ provides a useful relaxation of submodular maximization $f$ to continuous space, maintaining coordinate linearity and specific submodular properties. Its favorable properties have led to the application of multilinear extension in various prior works \citep{DBLP:journals/corr/BuchbinderF16,ene2016constrained,DBLP:journals/siamcomp/CalinescuCPV11,DBLP:journals/siamcomp/ChekuriVZ14} to design continuous methods for constrained submodular maximization, which typically involve two stages.
%
In the first stage, it starts with an empty solution $\mbx(0)$ and updates it during the time interval $[0, 1]$ by 
Frank-Wolfe or continuous greedy method to obtain a fractional solution that approximately maximizes the extension $F$.
In the second stage, the fractional solution $\mbx(1)$ is rounded to a feasible integral solution.

\paragraph{Challenges in continuous methods for $k$-submodular maximization.} 
Designing rounding schemes to $k$-submodular functions is a straightforward process (Lemma~\ref{lmm:rounding}). 
Therefore, we focus on the technical challenges of the first stage. We provide a summary of these challenges below. 
\begin{enumerate}
\item The domain of submodular extension is $[0,1]^n$, which benefits the closure of the coordinate-wise maximum operation, i.e., $\mbx\vee \mby\in [0,1]^n$ for all $\mbx, \mby\in [0,1]^n$. 
This property is crucial in the analysis of the approximation ratio of $\mbx(1)$. 
However, the domain of $k$-submodular extension is the corner of the cube $\Delta_k^n =\{\mbx\in[0,1]^{nk}: \sum\limits_{j=1}^k \mbx_{i,j}\leq 1,\forall i\in [n] \}$, rather than $[0,1]^{n k}$. Consequently, the closure property no longer holds.
\item Another beneficial property of submodular extension is the Lipschitzness \citep{DBLP:conf/aistats/BianMB017}, which regulates the value change of each step of continuous methods. 
However, it is unclear whether this property persists in $k$-submodular extension due to the additional structure of $k$ orthants.
\end{enumerate}

\paragraph{Technical idea: Auxiliary points and novel properties of our multilinear extension.}
To tackle the first challenge, we shift our focus to the linear combination operation, which involves constructing $a \mbx + (1-a) \mby$ ($a\in [0,1]$) for $\mbx, \mby\in \Delta_k^n$ instead of coordinate-wise maximum. 
This operation has the advantage of being closed in $\Delta_k^n$, inspiring us to construct auxiliary points for analysis. 
Specifically, given $\mathbf{o}^\star$ as the optimal fractional solution of $k$-submodular extension $F$ and $\mbx(t)\in t\cdot \Delta_k^n$ as the arriving point at time step $t\in [0,1]$, we create an auxiliary point $\mathbf{o}(t) = \mbx(t) + (1-t) \mathbf{o}^\star$, which is guaranteed to be in $\Delta_k^n$ due to the closure property. 
The use of auxiliary points has been previously explored in the literature \citep{DBLP:conf/soda/IwataTY16,DBLP:conf/nips/OhsakaY15,sakaue2017maximizing}, demonstrating to be useful for analysis purposes.

Then we investigate the relation between $F(\mbx(t))$ and $F(\mathbf{o}(t))$, whose key is to address the aforementioned second challenge.
When $f$ is monotone, we demonstrate that $F(\mbx(t+\delta)) - F(\mbx(t)) \gtrsim F(\mathbf{o}(t)) - F(\mathbf{o}(t + \delta))$, which directly leads to a conclusion that $F(\mbx(1))\gtrsim \frac{1}{2} F(\mathbf{o}^\star)$ (Lemma~\ref{lmm:frank}).
Theorem~\ref{thm:main_informal} is a direct corollary of this conclusion and the rounding guarantee (Lemma~\ref{lmm:rounding}).
We establish this result based on a novel observation of $k$-submodular extension, called \emph{approximate linearity} (Lemma~\ref{thm:properties}), which captures certain Lipschitzness of $F$ and allows us to estimate the increment $F(\mbx(t+\delta)) - F(\mbx(t))$ for sufficiently small values of $\delta$.

When $f$ is non-monotone, we require an additional property of $F$, called \emph{pairwise monotonicity} (Lemma~\ref{thm:properties}), to reduce the problem to the monotone case. Utilizing pairwise monotonicity, we are able to obtain an approximation $F(\mbx(1))\gtrsim \frac{1}{3} F(\mathbf{o}^\star)$ (Lemma~\ref{lmm:frank2}).
Similarly, Theorem~\ref{thm:nonmonotone_informal} is a direct corollary of this approximation and the rounding guarantee (Lemma~\ref{lmm:rounding}).

Overall, we discover novel properties of our $k$-submodular multilinear extension, including approximate linearity and pairwise monotonicity, which are useful in the analysis of auxiliary points.

\paragraph{Comparison with existing combinatorial approaches.}
We demonstrate that our approach using continuous optimization methods yields improved approximations for knapsack constraints compared to prior combinatorial methods such as those presented in~\citep{ha4395206improved,yu2023maximizing}. 
We offer intuitive explanations for this improvement and observe that a similar conclusion holds for submodular maximization with $O(1)$ knapsack constraints: to the best of our knowledge, no combinatorial method achieves an optimal approximation, whereas an optimal approximation algorithm via multilinear extension has been presented by \citep{DBLP:journals/siamcomp/ChekuriVZ14}. 
Our findings may suggest that the flexibility of continuous methods in selecting stepsizes and update directions provides an advantage over combinatorial approaches for handling knapsack constraints.





\subsection{Other related works}
\label{sec:related}

Submodular maximization, as a special case of $k$-submodular maximization, has a rich line of research with numerous results.
In the monotone case, tight $(1-1/e)$-approximations have been proposed for various constraints, such as single matroid constraint and $O(1)$ knapsacks constraint~\citep{DBLP:journals/siamcomp/CalinescuCPV11,DBLP:journals/siamcomp/ChekuriVZ14,DBLP:conf/soda/KulikST09,nemhauser1978analysis}. Furthermore, additional results have been developed for more complicated constraints, including the intersection of matroids and exchange systems~\citep{feldman2011improved,DBLP:journals/mor/LeeSV10}.
In the non-monotone case, the best-known approximation ratio for the single matroid or $O(1)$ knapsack constraint is $0.385$~\citep{DBLP:journals/corr/BuchbinderF16} while the hardness of $0.478$ holds for single matroid~\citep{DBLP:conf/soda/GharanV11}.

\paragraph{Concurrent work.}
Recent developments in \( k \)-submodular maximization research have introduced new algorithms with varying approximation ratios. 
For single matroid constraints, the threshold-decreasing algorithm in \citet{niu2023fast} achieves a \(1/2\)-approximation ratio for monotone objectives and a \(1/3\)-approximation ratio for non-monotone cases. 
For single knapsack constraints, an alternative greedy algorithm with \(0.432\)- and \(0.317\)-approximation ratios for monotone and non-monotone objectives, respectively, is presented in \citep{xiao2023approximation}.
In comparison, our algorithms outperform these approaches by achieving better approximation ratios or allowing more general types of constraints. 
Our algorithms also offer greater flexibility across various constraints and achieve a tight \(1/2\) approximation ratio for monotone objectives and a \(1/3\) approximation ratio for non-monotone objectives with $O(1)$ knapsack constraints and single matroid constraints.



\section{Preliminaries}
\label{sec:pre}

In this section, we first define the constrained $k$-submodular maximization problem and then present the notion of conjunction constraints.
Let $[n]$ be the ground set.
Throughout this paper, we assume there exists a value oracle $\mathcal{O}_f$ that answers $f(\mathbf{s})$ for any query $\mathbf{s}\in \{0,\ldots,k\}^{n}$.

\paragraph{Constrained $k$-submodular maximization.}
We first present the following support constraints.

\begin{definition}[\bf{Support constraints}]
Given a convex down-closed polytope $\mathcal{P}\subseteq [0,1]^n$,\footnote{``Down-closed'' represents that for every $x,y\in [0,1]^n$ satisfying that $x_i\leq y_i$ for all $i\in [n]$, we have $x\in \mathcal{P}$ if $y\in \mathcal{P}$.} we say an integral solution $\mathbf{s} \in \{0,\ldots,k\}^n$ is consistent to $\mathcal{P}$, denoted as $\mathbf{s} \sim \mathcal{P}$, if the identity vector of its support set satisfies constraint $\mathcal{P}$, i.e., $1_{\mbox{supp}(\mathbf{s})}\in \mathcal{P}$.
\end{definition}

Such support constraints are widely studied in the literature~\citep{DBLP:conf/nips/OhsakaY15,sakaue2017maximizing} and have various applications, such as multiple topics influence maximization with finite budget~\citep{qian2017constrained,zhang2019budget}.

We are ready to define the $k$-submodular maximization problem with support constraints.

\begin{problem}[\bf{$k$-submodular maximization with support constraints}]
\label{prob}
Given a $k$-submodular function $f: \{0,\ldots,k\}^{n} \to \mathbb{R}_{\geq 0}$ together with a value oracle $\mathcal{O}_f$
and a support constraint $\mathcal{P}\subseteq [0,1]^n$, the goal is to find a vector $S\subseteq [n]$ with $\mathbf{s}\sim \mathcal{P}$ that maximizes $f(\mathbf{s})$.
\end{problem}

The most commonly used support constraints are matroid constraints and knapsack constraints; defined as follows.

\begin{definition}[\bf{Matroids and matroid constraints}]
\label{def:matroid}
A matroid is a pair $\mathcal{M} = ([n], \mathcal{I}_\mathcal{M})$ where $\mathcal{I}_\mathcal{M} \subseteq 2^n$, such that
1) $\forall B\in \mathcal{I}_\mathcal{M},~A\subset B~\Rightarrow~ A\in \mathcal{I}_\mathcal{M} $;
2) $\forall A,B\in \mathcal{I}_\mathcal{M},~ |A|<|B|~\Rightarrow~\exists x\in B\setminus A$ s.t. $A\cup \{x\}\in \mathcal{I}_\mathcal{M}$.
%
The matroid constraint is defined as 
$\mathcal{P}_\mathcal{M} := \mbox{conv}\{1_I:I \in \mathcal{I}_\mathcal{M} \}$, i.e., the convex hull of all identity vectors $1_I$.
\end{definition}

Problem \ref{prob} is called $k$-submodular maximization with a matroid constraint when $\mathcal{P}=\mathcal{P}_\mathcal{M}$.

\begin{definition}[\bf{Knapsack constraints}]
\label{def:knapsack}
Given a non-negative matrix $A\in \mathbb{R}_{\geq 0}^{l\times n}$ ($l,n\in \mathbb{N}_+$) and a budget vector $b\in \mathbb{R}_+^l$, 
we call
$\mathcal{P}_\mathcal{K} := \{\mbx\in[0,1]^n:A\mbx\leq b\}$ the intersection of $l$ knapsack constraints.
Specifically, $\mathcal{P}_\mathcal{K}$ is called a knapsack constraint when $l=1$.
\end{definition}

Problem \ref{prob} is called k-submodular maximization with knapsack constraints when $\mathcal{P}=\mathcal{P}_\mathcal{K}$.
We may also consider the intersection of multiple matroid constraints $\mathcal{P}_{\mathcal{M}_1}, \ldots, \mathcal{P}_{\mathcal{M}_a}$ and knapsack constraints $\mathcal{P}_{\mathcal{K}}$, in which 
$\mathcal{P} = \left(\mathop{\bigcap}\limits_{i\in [l]}\mathcal{P}_{\mathcal{M}_i}\right)\cap \mathcal{P}_\mathcal{K}.
$

\paragraph{Conjunction constraint and its membership oracle.}
For ease of defining multilinear extension (Section~\ref{sec:extension}), we present another way of encoding support constraint $\mathcal{P}$, called conjunction constraint. 
For preparation, we define the corner of cube in $[0,1]^{nk}$ as $\Delta_k^n :=\{\mbx\in[0,1]^{nk}: \sum\limits_{j=1}^k \mbx_{i,j}\leq 1,\forall i\in [n] \}$.
Note that $\Delta_k^n$ can be viewed as a (partition) matroid constraint in $[nk]$. 
%
%
%
%
%

\begin{definition}[\bf{Conjunction constraints}]
Given a support constraint $\mathcal{P}\subseteq [0,1]^n$, we define its corresponding \emph{conjunction constraint} $\mathcal{P}^c\subseteq  \Delta_k^n$ as 
\[\mathcal{P}^c =\bigg\{\mbx\in \Delta_k^n: \mathbf{s}=\bigg(\sum\limits_{j=1}^k\mbx_{1,j},\ldots,\sum\limits_{j=1}^k\mbx_{n,j}\bigg)\in \mathcal{P} \bigg\}.
\]
\end{definition}

For every $\mathbf{s}_i\in [0,1]$, we divide it into $k$ dimensions $\mathbf{x}_{i,1}, \ldots, \mathbf{x}_{i,k}\in [0,1]$ satisfying that $\sum\limits_{j=1}^k\mbx_{i,j} = \mathbf{s}_i$, i.e., all such vectors $(\mathbf{x}_{i,1}, \ldots, \mathbf{x}_{i,k})$ form a (scaled) simplex in $\Delta_k$.
Then $\mathcal{P}^c$ can be viewed as a conjunction of these simplexes.
By definition, we have the following claim.

\begin{claim}
Given a support constraint $\mathcal{P}$, the conjunction constraint $\mathcal{P}^c$ is convex and down-closed.
\end{claim}

We also define the membership oracle $\mathcal{O}_{\mathcal{P}^c}$ of $\mathcal{P}^c$, that is for any query $\mbx \in \Delta_k^n$, $\mathcal{O}_{\mathcal{P}^c}$ answers whether $\mbx \in \mathcal{P}^c$ or not.
The following lemma shows the existence of $\mathcal{O}_{\mathcal{P}^c}$, which is useful for our algorithms.

\begin{lemma}[\bf{Existence of membership oracle \citep{DBLP:journals/jct/Cunningham84}}]
Given a support constraint $\mathcal{P}$ as the intersection of matroid constraints and knapsack constraints,
there exists an efficient membership oracle $\mathcal{O}_{\mathcal{P}^c}$ of the conjunction constraint $\mathcal{P}^c$.
\end{lemma}
\section{Results for monotone $k$-submodular maximization}
\label{sec:theoretical}

In this section, we consider the case that the objective $k$-submodular function $f$ is monotone. 

\begin{theorem}[\bf{Main theorem \rom{1}}]
\label{the:main}
There exists a polynomial-time algorithm for monotone $k$-submodular maximization with support constraint $\mathcal{P}\subseteq [0,1]^n$ that achieves 

\begin{itemize}
    \item $(\frac{1}{2}-\varepsilon)$-approximate with calling $\mathcal{O}_f$ at most ${O}\left(\frac{k^2n^6\log\left(\frac{n}{\varepsilon\eta}\right)}{\varepsilon^3}\right)$ times, for any fixed $\varepsilon > 0$
    when $\mathcal{P}$ is a single matroid constraint;
    \item $(\frac{1}{2}-\varepsilon)$-approximate with calling $\mathcal{O}_f$ at most  $O(n^{\textup{poly}(\frac{1}{\varepsilon})}+\frac{k^2n^6\log\left(\frac{n}{\varepsilon\eta}\right)}{\varepsilon^3})$ times, for any fixed $\varepsilon > 0$
    when $\mathcal{P}$ is the intersection of $O(1)$ knapsack constraints;
    \item $(\frac{0.3}{b}-\varepsilon)$-approximate with calling $\mathcal{O}_f$ at most   $O(n^{\textup{poly}(\frac{1}{\varepsilon})}+\frac{k^2n^6\log\left(\frac{n}{\varepsilon\eta}\right)}{\varepsilon^3})$ times, for any fixed $\varepsilon > 0$
    when $\mathcal{P}$ is the intersection of $b$ matroid constraints and $O(1)$ knapsack constraints;
    %
\end{itemize}
with probability at least $1-\eta$.
\end{theorem}

Existing algorithms are all based on combinatorial methods \citep{ha4395206improved,DBLP:journals/corr/abs-2307-13996,yu2023maximizing}. Compared to previous methods, our approach attains a nearly tight approximation ratio of \( \frac{1}{2} - \varepsilon \) for a single matroid constraint and \( O(1) \) for a knapsack constraint. 
Furthermore, we achieve an improved approximation ratio for the intersection of \( b \) matroid constraints and \( O(1) \) knapsack constraints. 
We remark that our query complexity is usually larger than existing combinatorial methods, e.g., a \( 0.432 \)-approximate algorithm with a query complexity of \( O(k^9 n^{10}) \) for a single knapsack constraint \citep{DBLP:journals/corr/abs-2307-13996}. 
The focus of this paper is to achieve (asymptotically) optimal approximation algorithms, which has been an open problem.


The rest of the section introduces the techniques used to obtain the results. We develop the multilinear extension (Section \ref{sec:extension}) and a unified optimization framework thereof (Section \ref{sec:alg}).
We then present the proof of Theorem \ref{the:main} in Section \ref{sec:performance}.

\subsection{Multilinear extension for $k$-submodular functions}
\label{sec:extension}

We first propose the multilinear extension of $k$-submodular functions.

\begin{definition}[\bf{Multilinear extension of $k$-submodular functions}]
\label{def:mul}
Given a $k$-submodular function $f\colon \{0,\ldots,k\}^{n}\to \R_{\geq 0}$, we define its multilinear extension $F\colon \Delta_k^n \to \R_{\geq 0}$ to be
\begin{equation}
\label{eqn:multilinear}
F(\mbx) = \sum\limits_{\mathbf{s}\in \{0,\ldots,k\}^{n}} f(\mathbf{s})  \prod\limits_{i\in [n]: \mathbf{s}_i \neq 0} \mbx_{i,\mathbf{s}_i} \prod\limits_{i\in [n]: \mathbf{s}_i = 0}\Big(1-\sum\limits_{j=1}^k\mbx_{i,j}\Big)\,.
\end{equation}
\end{definition}

The proposed definition is a natural generalization of the well-known multilinear extension of submodular functions ($k=1$)~\citep{DBLP:journals/siamcomp/CalinescuCPV11}. 
The domain $\Delta_k^n$ can be regarded as the extension of $[0,1]^n$ for submodular when $k=1$.
For every $\mbx\in \Delta_k^n$, it follows that $F(\mbx) = \EE[f(\mathbf{s})]$ where $\mathbf{s}\in \left\{0,\ldots,k\right\}^n$ denotes a random vector: for each item $i\in [n]$, $\mathbf{s}_i = j$ for $j\in [k]$ with a probability $\mbx_{i,j}$ and otherwise, $\mathbf{s}_i = 0$, which occurs independently across all items.
The following lemma presents good properties for our multilinear extension, which is useful for algorithm design.
The proof can be found in Appendix~\ref{app:property}.

\begin{lemma}[\bf{Properties of the  multilinear extension of $k$ submodular function}]
\label{thm:properties}
Let $f\colon \{0,\ldots,k\}^{n}\to \R_{\geq 0}$ be a $k$-submodular function. Then its multilinear extension $F\colon \Delta_k^n \to \R_{\geq 0}$ satisfies the following properties.
\begin{itemize}
    \item (Multilinearity) For every $i\in [n]$, $j\in [k]$ and $\mbx,\mbx'\in \Delta_k^n$ with $\mbx'-\mbx = c\cdot e_{i,j}$,\footnote{$e_{i,j}$ is the $(i,j)$-th unit basis vector in $\R^{n\times k}$.} the equality $\partial_{i,j} F(\mbx) = \partial_{i,j} F(\mbx')$ holds. 
    \item (Element-wise non-positive Hessian) 
    Let $M := \max \{\max_{i,j}F(\mathbf{e}_{i,j}) - F(\mathbf{0}),0\}$ be constant determined by $f$.
    For all  $ i_1,i_2\in [n], j_1,j_2\in [k]$,
\begin{equation*}
\frac{\partial^2 F}{\partial \mbx_{i_1,j_1}\partial \mbx_{i_2,j_2}}\left\lbrace
\begin{array}{cl}
 = 0 &\textup{ if } i_1 = i_2 ,           \\
{\in [-2M,0]} &\textup{ if } i_1 \neq i_2.
\end{array}\right. \, 
\end{equation*}
    \item (Pairwise monotonicity) For all $ i\in [n], j_1,j_2\in [k]$, $
\frac{\partial F}{\partial \mbx_{i,j_1}} + \frac{\partial F}{\partial \mbx_{i,j_2}}\geq 0$.
    \item (Approximate linearity) 
    For any points $\mbx, \mbx' \in \Delta_k^n$ satisfy that 
    $\mbx'-\mbx\in \delta\cdot \Delta_k^n$, then 
    \[F(\mbx')-F(\mbx)\geq ~\sum_{i\in[n],j\in [k] } (\mbx'_{i,j} -\mbx_{i,j} )\cdot \partial_{i,j} F(\mbx)~-n^2\delta^2M.\,\]
    \item (Preservation of monotonicity)  If $f$ is monotone, $F$ is monotone, i.e., for any point $\mbx \in \Delta_k^n$, $\partial_{i,j} F(\mbx)\geq 0$ for all $i\in [n]$ and $j\in [k]$.
\end{itemize}
\end{lemma}

As a generalization of submodular functions, the multilinear extension of $k$-submodular functions also exhibits multilinearity and non-positive Hessian elements. 
Furthermore, monotonicity is preserved by the extension.

Several novel properties emerge due to the inherent partition property of $k$-submodular functions. 
%
%
First, the Hessian of our extension contains zero-value elements in the same $i$'s blocks, i.e. $\partial^2 F/\partial \mbx_{i_1,j_1}\partial \mbx_{i_2,j_2} =0$ if $i_1 = i_2$, which is useful in designing rounding schemes. 
%
%
Our extension also exhibits an exclusive pairwise monotone property, which allows us to handle the non-monotone case.
More importantly, we demonstrate a novel approximate linearity property for the $k$-submodular case, which allows us to estimate the increment of movement with a sufficient small stepsize in the analyses of the Frank-Wolfe type of methods~\citep{DBLP:conf/aistats/BianMB017}. 
This property is analogous to the Lipschitz assumption in DR-submodular maximization \citep{DBLP:conf/aistats/BianMB017}.

%
%
%

We will need a (stochastic) gradient oracle \( \mathcal{O}_{\nabla F}^{ (\varepsilon,\eta) } \) for \( \nabla F \)
with parameters $\varepsilon,\delta\in (0,1)$ where for any query \( \mathbf{x} \in \Delta_k^n \), \( \mathcal{O}_{\nabla F}^{ {(\varepsilon,\eta})} \) provides
a stochastic estimate \( \widehat{\nabla F(\mathbf{x})} \) that is ``$\frac{\varepsilon}{k n^2}$-close'' to the gradient \( \nabla F(\mathbf{x}) \) in terms of $\ell_\infty$-norms, with a probability at least $1-\eta$.
We summarize this oracle in the following lemma, whose proof can be found in Appendix \ref{app:cher}.
\begin{lemma}[\bf{Existence of oracle \( \mathcal{O}_{\nabla F}^{{ (\varepsilon,\eta)}}\)}]
\label{lmm:concen}
Given $\varepsilon, \eta \in (0,1)$,
{let $F$ be the multilinear extension of $f$. 
There is an algorithm that for any point $\mbx\in \Delta_k^n$, {calls $\mathcal{O}_f$} for at most $\lceil\frac{16k^2n^6M^2\log\left( \frac{Mn^2+1}{\varepsilon\eta}\right)}{\varepsilon^2}\rceil$ times and returns a stochastic estimate \( \widehat{\nabla F(\mathbf{x})} \) of the gradient \( \nabla F(\mathbf{x}) \) such that for all $i\in [n]$ and $j\in [k]$, 
\[\left|\widehat{\partial_{i,j} F(\mathbf{x})}- \partial_{i,j} F(\mathbf{x})\right|\leq \frac{\varepsilon}{kn^2},
\]
with probability at least $1 - \frac{\varepsilon\eta}{Mn^2+1}$.}
\end{lemma}

As a direct corollary, we know that $\left\|\widehat{\nabla F(\mathbf{x})}- {\nabla F(\mathbf{x})} \right\|_2\leq \frac{\varepsilon}{n}$ holds for any point $\mbx\in \Delta_k^n$, which is useful for our analysis.

%

\subsection{Our algorithm}
\label{sec:alg}
In this section, we propose our main algorithm (Algorithm \ref{alg:main}) consisting of two stages: 1) using the Frank-Wolfe method to approximately maximize the multilinear extension (Lines 2-4); 2) rounding the fractional solution to an integral solution (Line 5).
Recall that $M = \max\left\{\max_{i,j} F(\mathbf{e}_{i,j}) - F(\mathbf{0}),0 \right\}$.

\begin{algorithm}[htp]
 \caption{Frank-Wolfe algorithm for monotone $k$-submodular maximization}
  \label{alg:main}
\DontPrintSemicolon
\SetNoFillComment
  \SetKwInOut{Input}{Input}
  \Input{{Parameters $\varepsilon, \eta\in (0,1)$;} oracles $\mathcal{O}_f$, $\mathcal{O}_{\nabla F}^{{(\varepsilon,\eta)}}$, $\mathcal{O}_{\mathcal{P}^c}$.}
    {\bfseries Initialize:} $\mathbf{x}(0) \leftarrow \mathbf{0}$, $t\leftarrow 0$; stepsize $\delta = \frac{1}{N}$ with  {$N = \lceil\frac{Mn^2}{\varepsilon}\rceil$}.  \; 
\While{$t<1$}{
 find a direction $\mathbf{v}(t) =  \mathop{\arg\max}\limits_{\mathbf{v}\in \mathcal{P}^c}\langle \widehat{\nabla F(\mathbf{x}(t))}, \mathbf{v}\rangle$ \Comment*[r]{By LP}
  $\mathbf{x}(t+\delta) = \mathbf{x}(t)+ \delta \mathbf{v}(t)$, $t\leftarrow t+\delta$\;
   }
 $\mathbf{s}\leftarrow \texttt{KSUBROUND}(\mathbf{x}(1))$. \Comment*[r]{by Lemma \ref{lmm:rounding}}
\Return $\mathbf{s}$
\end{algorithm}
%
In the first stage, the Frank-Wolfe algorithm stops at the $N$-th iteration.
In each iteration, we utilize the surrogate function $\langle \nabla F(\mathbf{s}(t)), \mathbf{v}(t)\rangle $ by searching for the feasible direction that maximizes the improvement in the function value. 
Finding such a direction amounts to maximizing a linear objective subject to a polytope in the positive orthant, and 
costs approximately the same as solving a positive LP, for which a nearly-linear time solver exists \cite{DBLP:conf/aistats/BianMB017, DBLP:conf/stoc/ZhuO15}.
In the second stage, the performance of our rounding scheme is presented in the following lemma, whose proof can be found in Appendix \ref{sec:rounding}. 
We remark that the rounding scheme works even for the non-monotone case, and hence, can also be applied in Section~\ref{sec:non-monotone}.

\begin{lemma}[\bf{Rounding scheme}]
\label{lmm:rounding}
Given a non-monotone $k$-submodular function $f$, its multilinear extension $F$, a support constraint $\mathcal{P}$ and {its relaxation  $\mathcal{P}^c$},
there is an algorithm $\mathtt{KSUBROUND(\cdot)}$ which runs in polynomial time and maps the fractional solution $\mathbf{x}\in \mathcal{P}^c$ to integral solution $\mathbf{s}\sim \mathcal{P}$ such that 
\begin{itemize}
    \item $\mathbb{E}\left[f(\mathbf{s})\right]\geq F(\mathbf{x})$ {without calling  $O_f$}, when $\mathcal{P}$ is single matroid constraint;\footnote{The rounding scheme for single matroid mainly applies the randomized swap rounding approach~\cite{DBLP:conf/focs/ChekuriVZ10,DBLP:journals/siamcomp/ChekuriVZ14}, which only needs to call membership oracles $\mathcal{O}_{\mathcal{P}^c}$ instead of $\mathcal{O}_f$.}
    \item $\mathbb{E}\left[f(\mathbf{s})\right]\geq (1-\varepsilon)F(\mathbf{x})$ for any fixed $\varepsilon > 0$  {with calling $O_f$ at most $O(n^{poly(1/\varepsilon}))$ times}, when $\mathcal{P}$ is $l=O(1)$ knapsack constraints;
    \item $\mathbb{E}\left[f(\mathbf{s})\right]\geq \left(\frac{0.6}{b}F(\mathbf{x})-\varepsilon\right)$  {for any fixed $\varepsilon > 0$ with calling $O_f$ at most $O(n^{poly(1/\varepsilon}))$ times}, when $\mathcal{P}$ is the intersection of $b$ matroid constraints and $l=O(1)$  knapsack constraints.
\end{itemize}
\end{lemma}


It is worth notice that the previous combinatorial algorithms \citep{ha4395206improved,DBLP:conf/soda/IwataTY16,DBLP:conf/nips/OhsakaY15,sakaue2017maximizing,tang2022maximizing,DBLP:journals/talg/WardZ16,yu2023maximizing} can be viewed as a modified Frank-Wolfe algorithm that iteratively moves along one coordinate with the stepsize equal to $1$. 
Our algorithm, on the other hand, can move along a flexible direction with a flexible stepsize, which enables us to achieve better approximations for knapsack constraints.

\subsection{Proof of Theorem \ref{the:main}: Performance analysis of Algorithm \ref{alg:main}}
\label{sec:performance}

It suffices to prove the following key lemma.
By the selection of $\delta$ in Algorithm~\ref{alg:main}, Theorem~\ref{the:main} is a direct corollary of Lemmas \ref{lmm:rounding} and \ref{lmm:frank}.

\begin{lemma}[\bf{Analysis of the Frank-Wolfe algorithm}]
\label{lmm:frank}
Let $\mathbf{o}^\star = \mathop{\arg\max}\limits_{\mbx\in\mathcal{P}^c}F(\mbx)$. Then 
$F(\mathbf{x}(1))\geq \frac{1}{2} F(\mathbf{o}^\star)-
\varepsilon$, with probability at least $1-\eta$.
\end{lemma}

Here, we assume $M$ is constant by re-scaling $f$.
This is because we can require that $F(\mathbf{x}(1))\geq \frac{1}{2} F(\mathbf{o}^\star)-\varepsilon M$ in the above lemma, which implies Theorem~\ref{the:main}. 
The key idea of Lemma~\ref{lmm:frank} is to analyze the value gain of each iteration.
Following the commonly used idea to $k$-submodular maximization~\citep{DBLP:conf/soda/IwataTY16,DBLP:conf/nips/OhsakaY15,sakaue2017maximizing}, we construct an auxiliary sequence $\mathbf{o}(t) = \mbx(t)+(1-t)\mathbf{o}^\star$ to be a linear combination of $\mathbf{o}^\star$ and $\mbx(t)$ such that $\mathbf{o}(t)$ is still contained in $\mathcal{P}^c$.
Such sequence satisfies that $\mathbf{o}(0) =\mathbf{o}^\star $ and $\mathbf{o}(1) = \mathbf{x}(1)$. 
Then it suffices to compare the decrease of the auxiliary sequence $F(\mathbf{o}(t))-F(\mathbf{o}(t+\delta))$ and the increase of the solution sequence $F(\mathbf{x}(t+\delta))-F(\mathbf{x}(t))$.

\begin{proof}[Proof of Lemma~\ref{lmm:frank}]
To obtain the guarantee, we construct the following auxiliary sequence.
\[\mathbf{o}(t) = \mathbf{x}(t)+\left(1-t\right)\mathbf{o}^\star,\]
\[\mathbf{o}(t+\delta) = \mathbf{x}(t)+\delta \mathbf{v}(t) + \left(1-t-\delta\right)\mathbf{o}^\star,\]
\[\mathbf{o}'(t) = \mathbf{x}(t)+\left(1-t-\delta\right)\mathbf{o}^\star.\]
By induction on $t$ and the definition of $\mathbf{x}(t)$, we obtain $\frac{1}{t}\mathbf{x}(t) = \sum\limits_{i=1}^{t/\delta}\frac{\delta}{t}\mathbf{v}(i)$. Thus, $\frac{1}{t}\mathbf{x}(t)$ can be expressed as a linear combination of $\mathbf{v}(1),\ldots,\mathbf{v}(t)$. Since $\mathbf{v}(t)\in \mathcal{P}^c\subseteq \Delta_k^n$, it follows that $\mathbf{x}(t)\in t \cdot{\mathcal{P}^c}$, which implies that $\mathbf{o}(t), \mathbf{o}(t+\delta)\in \mathcal{P}^c$. 
{By the definition of $\mathbf{o}'(t)$ and $\mathbf{o}(t+\delta)$, we have 
\[\mathbf{o}(t+\delta)- \mathbf{o}'(t) = \mathbf{v}(t)\in \mathcal{P}^c\subseteq[0,1]^{nk}.\]
Combining the monotonicity of $F$, we have 
\begin{equation}
    \label{eq:3}
F(\mathbf{o}'(t))-F(\mathbf{o}(t+\delta))\leq 0.
\end{equation}
}
Now we bound the improvement in every step.
{\begin{align*}
&F(\mathbf{o}(t))-F(\mathbf{o}(t+\delta))\\ =~&  F(\mathbf{o}(t))-F(\mathbf{o}'(t))+F(\mathbf{o}'(t))-F(\mathbf{o}(t+\delta)) \\
\leq ~&  F(\mathbf{o}(t))-F(\mathbf{o}'(t))\tag*{(by Eq.~\eqref{eq:3})} \\
\leq ~& \langle \nabla F(\mathbf{o}'(t)), \mathbf{o}^\star\rangle \delta \tag*{(by submodularity and multilinearity)}\\
= ~& \sum\limits_{i,j} \partial_{i,j} F(\mathbf{o}'(t)) \mathbf{o}^\star_{i,j}\delta \\
\leq ~& \sum\limits_{i,j} \partial_{i,j} F(\mathbf{x}(t)) \mathbf{o}^\star_{i,j}\delta\tag*{(by submodularity and $\mathbf{o}^\star_{i,j}\geq 0$)}  \\
= ~& \langle \nabla F(\mathbf{x}(t)), \mathbf{o}^\star\rangle \delta \\
= ~& \langle \widehat{\nabla F(\mathbf{x}(t))}, \mathbf{o}^\star\rangle \delta + \langle \nabla F(\mathbf{x}(t)) -  \widehat{\nabla F(\mathbf{x}(t))}, \mathbf{o}^\star\rangle \delta \\
\leq ~&   \langle \widehat{\nabla F(\mathbf{x}(t))}, \mathbf{v}(t)\rangle \delta + \langle \nabla F(\mathbf{x}(t)) -  \widehat{\nabla F(\mathbf{x}(t))}, \mathbf{o}^\star\rangle \delta  \tag*{(by choice of $\mathbf{v}(t)$)} \\
= ~&   \langle {\nabla F(\mathbf{x}(t))}, \mathbf{v}(t)\rangle \delta + \langle \nabla F(\mathbf{x}(t)) -  \widehat{\nabla F(\mathbf{x}(t))}, \mathbf{o}^\star\rangle \delta  +  \langle \widehat{\nabla F(\mathbf{x}(t)) }-  {\nabla F(\mathbf{x}(t))},\mathbf{v}(t)\rangle \delta \\
\leq ~&\langle {\nabla F(\mathbf{x}(t))}, \mathbf{v}(t)\rangle \delta + \left\|\nabla F(\mathbf{x}(t)) -  \widehat{\nabla F(\mathbf{x}(t))}\right\|_2 \left\|\mathbf{o}^\star\right\|_2 \delta\\
~&+ \left\|\nabla F(\mathbf{x}(t)) -  \widehat{\nabla F(\mathbf{x}(t))}\right\|_2 \left\|\mathbf{v}(t)\right\|_2 \delta\tag*{(by Cauchy–Schwarz inequality)}  \\
\leq ~&\langle {\nabla F(\mathbf{x}(t))}, \mathbf{v}(t)\rangle \delta + \frac{\varepsilon}{n} \left\|\mathbf{o}^\star\right\|_2 \delta + \frac{\varepsilon}{n} \left\|\mathbf{v}(t)\right\|_2 \delta\tag*{(by Lemma \ref{lmm:concen})} \\
\leq~& F(\mathbf{x}(t+\delta))- F(\mathbf{x}(t)) + n^2\delta^2M+\varepsilon\delta.\tag*{(by approximate linearity)}\\
\leq~& F(\mathbf{x}(t+\delta))- F(\mathbf{x}(t)) + 2\varepsilon\delta.\tag*{(by choice of $\delta$)}
\end{align*}}

By Lemma \ref{lmm:concen}, the above inequality holds with probability at least $1-\frac{\varepsilon\eta}{Mn^2+1}$. 
Thus, by union bound over $N = \lceil\frac{Mn^2}{\varepsilon}\rceil$ steps, we conclude that
$
F(\mathbf{o}(0))-F(\mathbf{o}(1))\leq F(\mathbf{x}(1))- F(\mathbf{x}(0))+  2\varepsilon 
$ holds with probability at least  $1-\eta$.
\end{proof}

{Finally, we analyze the query complexity of Algorithm \ref{alg:main}. In Lines 2-4, Algorithm \ref{alg:main} queries \( \mathcal{O}_{\nabla F} \) a total of \( N = \lceil \frac{Mn^2}{\varepsilon} \rceil \) times. This implies that the query complexity with respect to \( f \) is bounded as
\[\# \mbox{Calls to~} \mathcal{O}_f \leq \lceil\frac{Mn^2}{\varepsilon}\rceil\cdot\lceil\frac{16k^2n^6M^2\log\left( \frac{Mn^2+1}{\varepsilon\eta}\right)}{\varepsilon^2}\rceil = {O}\left(\frac{k^2n^6\log\left(\frac{n}{\varepsilon\eta}\right)}{\varepsilon^3}\right).\]
Combining Lemma \ref{lmm:rounding}, Theorem \ref{the:main} follows.
}


\section{Results for non-monotone $k$-submodular maximization}
\label{sec:non-monotone}

In this section, we present an algorithm (Algorithm~\ref{alg:nonmonotone}) and its analysis (Theorem~\ref{the:nonmonotone}) for the non-monotone $k$-submodular objective.
Recall that $M = \max\left\{\max_{i,j} F(\mathbf{e}_{i,j}) - F(\mathbf{0}) ,0\right\}$.

\begin{theorem}[\bf{Main theorem \rom{2}}]
\label{the:nonmonotone}
%
There exists a polynomial-time algorithm for non-monotone $k$-submodular maximization with support constraint $\mathcal{P}\subseteq [0,1]^n$  that achieves 
\begin{itemize}
    \item {$(\frac{1}{3}-\varepsilon)$-approximation and calls $\mathcal{O}_f $ at most ${O}\left(\frac{k^2n^6\log\left(\frac{n}{\varepsilon\eta}\right)}{\varepsilon^3}\right)$ times, for any fixed $\varepsilon > 0$
    when $\mathcal{P}$ is a single matroid constraint};
    \item {$(\frac{1}{3}-\varepsilon)$-approximate and calls $\mathcal{O}_f $ at most $O\left(n^{poly(1/\varepsilon)}+\frac{k^2n^6\log\left(\frac{n}{\varepsilon\eta}\right)}{\varepsilon^3}\right)$ times, for any fixed $\varepsilon > 0$
    when $\mathcal{P}$ is the intersection of $O(1)$ knapsack constraints};
    \item {$(\frac{0.2}{b}-\varepsilon)$-approximate and calls $\mathcal{O}_f $ at most $O\left((n^{poly(1/\varepsilon)}+\frac{k^2n^6\log\left(\frac{n}{\varepsilon\eta}\right)}{\varepsilon^3}\right)$ times, for any fixed $\varepsilon > 0$
    when $\mathcal{P}$ is the intersection of $b$ matroid constraints and $O(1)$ knapsack constraints};
\end{itemize}
{with probability at least $1-\eta$.}
\end{theorem}

%
%

\begin{algorithm}[H]
 \caption{Frank-Wolfe algorithm for non-monotone $k$-submodular maximization}
  \label{alg:nonmonotone}
\DontPrintSemicolon
\SetNoFillComment
  \SetKwInOut{Input}{Input}
  \Input{{Parameters $\varepsilon, \eta\in (0,1)$;} oracles $\mathcal{O}_f$, $\mathcal{O}_{\nabla F}^{{(\varepsilon,\eta)}}$, $\mathcal{O}_{\mathcal{P}^c}$.}
    {\bfseries Initialize:} $\mathbf{x}(0) \leftarrow \mathbf{0}$, $t\leftarrow 0$; stepsize $\delta = \frac{1}{N}$ with $N = \lceil\frac{Mn^2}{\varepsilon}\rceil$.\; 
\While{$t<1$}{
 Find a direction $\mathbf{v}(t) =  \mathop{\arg\max}\limits_{\mathbf{v}\in \mathcal{P}^c}\langle \widehat{\nabla F(\mathbf{x}(t))}, \mathbf{v}\rangle$. \Comment*[r]{By LP}
 {\bfseries Initialize:} ${\mathbf{v}}'(t) = \mathbf{0}$.\;
 \For{$i\in \mbox{supp}(\mathbf{v}(t))$}{
Order partial derivative as
 $\widehat{\partial_{i,j_1}F(\mathbf{x}(t))} \geq \widehat{\partial_{i,j_2}F(\mathbf{x}(t))}\geq \ldots \geq  \widehat{\partial_{i,j_k}F(\mathbf{x}(t))}$.\;
 \If{$\widehat{\partial_{i,j_2}F(\mathbf{x}(t))}\geq 0$}{
 ${\mathbf{v}}'_{i,j_2}(t) \leftarrow \sum_{j\in [k]}{v_{i,j}(t)}$.
 }
 }
  $\widehat{\mathbf{v}}(t) = \frac{1}{2}\left({\mathbf{v}}(t) + {\mathbf{v}}'(t)\right)$.\;
  $\mathbf{x}(t+\delta) = \mathbf{x}(t)+ \delta \widehat{\mathbf{v}}(t)$, $t\leftarrow t+\delta$.\;
   }
 $\mathbf{s}\leftarrow \texttt{KSUBROUND}(\mathbf{x}(1))$. \Comment*[r]{by Lemma \ref{lmm:rounding}}
\Return $\mathbf{s}$
\end{algorithm}

Similar to Algorithm \ref{alg:main}, Algorithm~\ref{alg:nonmonotone} also contains two stages: A Frank-Wolfe-type method that computes a fraction solution $\mbx(1) \in \mathcal{P}^c$ (Lines 2-10) and a rounding procedure (Line 11).
The main difference is in the first stage, where Algorithm \ref{alg:nonmonotone} moves along the complemented direction $\widehat{\mathbf{v}}(t)$ as an average of the locally optimal direction ${\mathbf{v}}(t)$ and vector $v'(t)$ depending on the signal of the second largest partial derivatives $\partial_{i,j_2}F(\mathbf{x}(t))$ for every $i\in [n]$.
This construction is motivated by the pairwise monotonicity of $F$, which enables us to reduce the non-monotone case to the monotone one in the analysis. 
Now we prove Theorem \ref{the:nonmonotone}.
Similar to Lemma~\ref{lmm:frank}, we first summarize the quality of the fractional solution $\mbx(1)$ in the following lemma.
\begin{lemma}
\label{lmm:frank2}
Let $\mathbf{o}^\star = \mathop{\arg\max}\limits_{\mbx\in  \mathcal{P}^c}F(\mbx)$. Then 
$F(\mathbf{x}(1))\geq \frac{1}{3} F(\mathbf{o}^\star)-2\varepsilon$, with probability at least $1 - \eta$.
\end{lemma}
%
%
%
%
\begin{proof}
To obtain the guarantee, we construct the following auxiliary sequence:
\[\mathbf{o}(t) = \mathbf{x}(t)+\left(1-t\right)\mathbf{o}^\star,\]
\[\mathbf{o}(t+\delta) = \mathbf{x}(t)+\delta \widehat{\mathbf{v}}(t) + \left(1-t-\delta\right)\mathbf{o}^\star,\]
\[\mathbf{o}'(t) = \mathbf{x}(t)+\left(1-t-\delta\right)\mathbf{o}^\star.\]
%
%
Now we consider a fixed $i\in [n]$ in Line 6.
Note that the feasibility of support constraint $\mathcal{P}^c$ is only affected by $\sum_{j\in [k]} \mathbf{v}_{i,j}$.
Hence, we have $\mathbf{v}_{i,j}(t) = \left\{\begin{array}{cc}
   \sum\limits_{j'\in [k]}\mathbf{v}_{i,j'}(t)  & j=j_1, \\
  0  & j\neq j_1,
\end{array}\right.$ by the definition of $j_1$.\footnote{Given the equivalency of the support constraint for all $\mathbf{v}_{i,j}(t)$, there should only be one unique non-zero value $j_!$, ensuring the auxiliary linear function achieves its maximum.}
Next, we discuss two cases based on the signal of $\partial_{i,j_2}F(\mathbf{x}(t))$ {at each step.
We remind the concentration property of the gradient estimators: for all $i\in [n]$ and $j\in [k]$, 
\begin{equation}
    \label{eq:4}
\left|\widehat{\partial_{i,j} F(\mathbf{x})}- \partial_{i,j} F(\mathbf{x})\right|\leq \frac{\varepsilon}{kn^2}~~~\mbox{    and 
}~~~\left\|\widehat{\nabla F(\mathbf{x})}- {\nabla F(\mathbf{x})} \right\|_2\leq \frac{\varepsilon}{n}
\end{equation}
with probability at least $1 - \frac{\varepsilon\eta}{Mn^2+1}$.}

\paragraph{Case 1: $\widehat{\partial_{i,j_2}F(\mathbf{x}(t))} < 0$.} 
By submodularity, we have $\partial_{i,j_2}F(\mathbf{o}'(t)) \leq  \partial_{i,j_2}F(\mathbf{x}(t))< 0$.
{Combining Eq. \eqref{eq:4}, we have $\widehat{\partial_{i,j_2}F(\mathbf{o}'(t)) }\leq \partial_{i,j_2}F(\mathbf{o}'(t)) + \frac{\varepsilon}{kn^2}\leq  \widehat{\partial_{i,j_2}F(\mathbf{x}(t))}+\frac{2\varepsilon}{kn^2}< \frac{2\varepsilon}{kn^2}.$}
By pairwise monotonicity (Lemma~\ref{thm:properties}), we have 
$\partial_{i,j_1}F(\mathbf{o}'(t))+ \partial_{i,j_2}F(\mathbf{o}'(t)) \geq 0$,
{Combining Eq. \eqref{eq:4}, we have $\widehat{\partial_{i,j_1}F(\mathbf{o}'(t))} + \widehat{\partial_{i,j_2}F(\mathbf{o}'(t))}\geq-\frac{2\varepsilon}{kn^2}.$}
Thus $\partial_{i,j_1}F(\mathbf{o}'(t))\geq {-\frac{4\varepsilon}{kn^2}}$.
Thus, we have 
\[\langle \nabla_i F(\mathbf{o}'(t)), \widehat{\mathbf{v}}_i(t)\rangle  \geq   {-\frac{4\varepsilon}{kn^2}}.\]
Moreover, due to the fact that $\widehat{\mathbf{v}}_i(t) = \frac{1}{2}\mathbf{v}_i(t)$, we have 
\[\langle \nabla_i F(\mathbf{x}(t)), \mathbf{v}_i(t)\rangle = 2\langle \nabla_i F(\mathbf{x}(t)), \widehat{\mathbf{v}}_i(t)\rangle.\]
\paragraph{Case 2: $\widehat{\partial_{i,j_2}F(\mathbf{x}(t)) \geq 0}$.}
By the definition of $\widehat{\mathbf{v}}(t)$ and $\mathbf{v}'(t)$,
we have $$2\widehat{\mathbf{v}}_{i,j}(t) = {\mathbf{v}}_{i,j}(t) + {\mathbf{v}}'_{i,j}(t) = \left\{\begin{array}{cc}
   \sum\limits_{j'\in [k]}\mathbf{v}_{i,j'}(t)  & j=j_1, \\
  \sum\limits_{j'\in [k]}\mathbf{v}_{i,j'}(t)  & j= j_2,\\
  0 &\mbox{otherwise.}
\end{array}\right.$$
Combining the pairwise monotonicity that $\partial_{i,j_1}F(\mathbf{o}'(t))+ \partial_{i,j_2}F(\mathbf{o}'(t)) \geq 0$,
  we have 
\[\langle \nabla_i F(\mathbf{o}'(t)), 
\widehat{\mathbf{v}}_i(t)\rangle \geq 0.\]
%
%
{By the condition that $\widehat{\partial_{i,j_2}F(\mathbf{x}(t)) \geq 0}$ and Eq. \eqref{eq:4}, we have $\partial_{i,j_2}F(\mathbf{x}(t)) \geq -\frac{\varepsilon}{kn^2}$, which implies that $\langle \widehat{\nabla_i F(\mathbf{x}(t))}, {\mathbf{v}}'_i(t)\rangle\geq -\frac{\varepsilon}{kn^2}$.} 
Combining the fact that $\widehat{\mathbf{v}}_i(t) = \frac{1}{2}\mathbf{v}_i(t)+\frac{1}{2}\mathbf{v}'_i(t)$, we have 
\begin{align*}
& \quad \langle \nabla_i F(\mathbf{x}(t)), {\mathbf{v}}_i(t)\rangle\leq \langle \nabla_i F(\mathbf{x}(t)), {\mathbf{v}}_i(t)\rangle + \langle \nabla_i F(\mathbf{x}(t)), {\mathbf{v}}'_i(t)\rangle + {\frac{\varepsilon}{kn^2}} \\
= & \quad 2\langle \nabla_i F(\mathbf{x}(t)), \widehat{\mathbf{v}}_i(t)\rangle+ {\frac{\varepsilon}{kn^2}}.
\end{align*}

Combining these two cases and the approximate linearity, we have 
\begin{equation}
    \label{eq:non1}
F(\mathbf{o}'(t))-F(\mathbf{o}(t+\delta))  \leq -\langle \nabla F(\mathbf{o}'(t)), \widehat{\mathbf{v}}(t)\rangle \delta +  n^2\delta^2M \leq  {\frac{4\varepsilon\delta}{kn}+n^2\delta^2M},
\end{equation}
and 
\begin{equation}
    \label{eq:non2}
\langle \nabla F(\mathbf{x}(t)), {\mathbf{v}}(t)\rangle \leq 2\langle \nabla F(\mathbf{x}(t)), \widehat{\mathbf{v}}(t)\rangle+ {\frac{\varepsilon}{kn}}.
\end{equation}

By definition $\mathbf{o}(1) = \mathbf{x}(1)$ and $\mathbf{o}(0) = \mathbf{o}^\star$. We bound the improvement in every step by the following inequalities.
\begin{align*}
&F(\mathbf{o}(t))-F(\mathbf{o}(t+\delta))\\ =~&  F(\mathbf{o}(t))-F(\mathbf{o}'(t))+F(\mathbf{o}'(t))-F(\mathbf{o}(t+\delta)) \\
\leq ~&  F(\mathbf{o}(t))-F(\mathbf{o}'(t)) +\frac{4\varepsilon\delta}{kn}+n^2\delta^2M\tag*{(by Eq.~\eqref{eq:non1})} \\
\leq ~& \langle \nabla F(\mathbf{o}'(t)), \mathbf{o}^\star\rangle \delta + \frac{4\varepsilon\delta}{kn}+n^2\delta^2M\tag*{(by submodularity and multilinearity)}\\
\leq ~& \langle \nabla F(\mathbf{x}(t)), \mathbf{o}^\star\rangle \delta + \frac{4\varepsilon\delta}{kn}+n^2\delta^2M\tag*{(by submodularity)}\\
= ~& \langle \widehat{\nabla F(\mathbf{x}(t))}, \mathbf{o}^\star\rangle \delta + \langle \nabla F(\mathbf{x}(t)) -  \widehat{\nabla F(\mathbf{x}(t))}, \mathbf{o}^\star\rangle \delta  + \frac{4\varepsilon\delta}{kn}+n^2\delta^2M\\
\leq ~&   \langle \widehat{\nabla F(\mathbf{x}(t))}, \mathbf{v}(t)\rangle \delta + \langle \nabla F(\mathbf{x}(t)) -  \widehat{\nabla F(\mathbf{x}(t))}, \mathbf{o}^\star\rangle \delta + \frac{4\varepsilon\delta}{kn}+n^2\delta^2M \tag*{(by choice of $\mathbf{v}(t)$)} \\
= ~&   \langle {\nabla F(\mathbf{x}(t))}, \mathbf{v}(t)\rangle \delta + \langle \nabla F(\mathbf{x}(t)) -  \widehat{\nabla F(\mathbf{x}(t))}, \mathbf{o}^\star\rangle \delta  +  \langle \widehat{\nabla F(\mathbf{x}(t)) }-  {\nabla F(\mathbf{x}(t))},\mathbf{v}(t)\rangle \delta\\
& ~ + \frac{4\varepsilon\delta}{kn}+n^2\delta^2M\\
\leq ~&\langle {\nabla F(\mathbf{x}(t))}, \mathbf{v}(t)\rangle \delta + \left\|\nabla F(\mathbf{x}(t)) -  \widehat{\nabla F(\mathbf{x}(t))}\right\|_2 \left\|\mathbf{o}^\star\right\|_2 \delta\\
&~ + \left\|\nabla F(\mathbf{x}(t)) -  \widehat{\nabla F(\mathbf{x}(t))}\right\|_2 \left\|\mathbf{v}(t)\right\|_2 \delta+ \frac{4\varepsilon\delta}{kn}+n^2\delta^2M \tag*{(by Cauchy–Schwarz inequality)}\\
\leq ~&\langle {\nabla F(\mathbf{x}(t))}, \mathbf{v}(t)\rangle \delta + \frac{\varepsilon}{n} \left\|\mathbf{o}^\star\right\|_2 \delta + \frac{\varepsilon}{n} \left\|\mathbf{v}(t)\right\|_2 \delta + \frac{4\varepsilon\delta}{kn}+n^2\delta^2M\tag*{(by Lemma \ref{lmm:concen})} \\
\leq ~&\langle \nabla F(\mathbf{x}(t)), \mathbf{v}(t)\rangle \delta+ 2\varepsilon\delta +  \frac{4\varepsilon\delta}{kn}+n^2\delta^2M\\
\leq ~&2\langle \nabla F(\mathbf{x}(t)), \widehat{\mathbf{v}}(t)\rangle \delta+  2\varepsilon\delta +  \frac{5\varepsilon\delta}{kn}+n^2\delta^2M\tag*{(by Eq.~\eqref{eq:non2})}\\
\leq~& 2\left(F(\mathbf{x}(t+\delta))- F(\mathbf{x}(t))\right) + 2\varepsilon\delta +  \frac{5\varepsilon\delta}{kn}+2n^2\delta^2M.\tag*{(by approximate linearity)}
\end{align*}
with probability at least $1 - \frac{\varepsilon\eta}{Mn^2+1}$.
{To sum the above inequalities over \( t = 0, \delta, \ldots, 1 \) and apply the union bound on the probability
}, we conclude that
\[
F(\mathbf{o}(0))-F(\mathbf{o}(1))\leq 2\left(F(\mathbf{x}(1))- F(\mathbf{x}(0))\right)+ 5\varepsilon, 
\]
with probability at least $1 - \eta$.
\end{proof}
Finally, the query complexity of Algorithm \ref{alg:nonmonotone} is identical to that of Algorithm \ref{alg:main}.
We complete the proof of Theorem \ref{the:nonmonotone} by combining Lemma \ref{lmm:frank2} and Lemma \ref{lmm:rounding}.

\section{Conclusions and future works}

We proposed unified Frank-Wolfe-type frameworks that solve $k$-submodular maximization with various settings. Notably, for single matroid constraint and $O(1)$ knapsacks constraint, we obtained an optimal $1/2$-approximation for monotone $k$-submodular functions and an optimal $1/3$-approximation for non-monotone $k$-submodular functions. Our frameworks also work for various constraint types including any combinations of matroid constraints and knapsack constraints.

Our frameworks are based on the multilinear extension of $k$-submodular functions. This provides a new way to design $k$-submodular maximization algorithms with flexible step sizes and flexible update directions. Considering that multilinear extension for submodular functions has obtained optimal results in many settings of submodular maximization, our extension and rounding techniques could be of independent interest for other $k$-submodular problems.


Many interesting directions are for further investigations. One question is whether $k$-submodular maximization algorithms can be derandomized. Such derandomized algorithms could benefit applications for better reproducibility and consistency. Another direction is to investigate optimization problems of pairwise monotone functions for non-monotone $k$-submodular maximization. Such functions are not well understood yet and our algorithm only uses this property in very simple ways.

\bibliography{references}
\bibliographystyle{plainnat}

\newpage

\appendix
\newpage
\section{Proof of Lemma \ref{lmm:concen}: Existence of an efficient oracle $\mathcal{O}_{F}$}
\label{app:cher}

{Given oracle access to a $k$-submodular function $f$, the Chernoff bounds~\citep{DBLP:books/cu/MotwaniR95} implies the following theorem which allows us to approximate the value of the multilinear extension $F$ to arbitrary accuracy.}

\begin{lemma}
\label{lmm:chernoff}
Assume $F$ is the multilinear extension of $f$. 
Given a point $\mbx \in  \mathcal{P}^c\subseteq \Delta_k^n$ and parameters $\varepsilon,\delta\in (0,1)$,
if $\mathbf{s}^1,\ldots, \mathbf{s}^t$ {are random vectors independently sampled as follows: for each $l\in [t]$, for each item $i\in [n]$, $\mathbf{s}_i^l = j$ for $j\in [k]$ with probability $\mbx_{i,j}$ and otherwise, $\mathbf{s}_i^l = 0$, which occurs independently across all items}; then { for any $\varepsilon_0\in (0,1)$, we have }
\[\left|\dfrac{1}{t}\sum\limits_{i=1}^t f(\mathbf{s}^i)-F(\mbx) \right|\leq\varepsilon_0|\max\limits_{\mathbf{s}\in \Delta_k^n}  f(\mathbf{s})|\]
with probability at least $1 - e^{-t\varepsilon_0^2/4}$.
\end{lemma}
{
For any partial derivative $\partial_{i,j}F(\mbx)$ at point $\mbx \in \Delta_k^n$ and direction $\mathbf{e}_{i,j}$, we construct its stochastic estimate $\widehat{\partial_{i,j} F(\mathbf{x})}$ as follows.
Consider points $\mbx^0, \mbx^1\in \Delta_k^n$ defined as
\[\mbx^0_{p,q} = \left\{\begin{array}{cc}
    0 & \mbox{If } p = i,\\
   \mbx_{p,q}  & \mbox{Otherwise}. 
\end{array}\right.\mbox{ and    }~~~~\mbx^1_{p,q} = \left\{\begin{array}{cc}
    0 & \mbox{If } p = i, q\neq j,\\
    1 & \mbox{If } p = i, q = j,\\
   \mbx_{p,q}  & \mbox{Otherwise}. 
\end{array}\right.\]
We observe that the Hessian elements of \( F \) satisfy the condition $
\frac{\partial^2 F}{\partial x_{i,j_1} \partial x_{i,j_2}} = 0$,
for all \( i \in [n] \) and \( j_1, j_2 \in [k] \). This implies that \( \partial_{i,j} F(\mathbf{x}^0) = \partial_{i,j} F(\mathbf{x}) \). Leveraging the multilinearity of \( F \), we deduce that 
\[ F(\mathbf{x}^1) - F(\mathbf{x}^0) = \partial_{i,j} F(\mathbf{x}^0) = \partial_{i,j} F(\mathbf{x}) .\]
{
We consider two sets of independent samples of random vectors, \( \mathbf{s}^{0,1}, \ldots, \mathbf{s}^{0,t} \) and \( \mathbf{s}^{1,1}, \ldots, \mathbf{s}^{1,t} \), which satisfy the property delineated in Lemma \ref{lmm:chernoff} for the points \( \mathbf{x}^0 \) and \( \mathbf{x}^1 \), respectively.
}
Define $\widehat{\partial_{i,j} F(\mathbf{x})}  = \dfrac{1}{t}\sum\limits_{i=1}^t f(\mathbf{s}^1_i)-\dfrac{1}{t}\sum\limits_{i=1}^t f(\mathbf{s}^0_i)$. Then by Lemma \ref{lmm:chernoff} the concentration property holds as
\begin{align*}
 \left|\widehat{\partial_{i,j} F(\mathbf{x})}- \partial_{i,j} F(\mathbf{x})\right|&~=~   \left|\dfrac{1}{t}\sum\limits_{i=1}^t f(\mathbf{s}^1_i)-F(\mathbf{x}^1) -\left(\dfrac{1}{t}\sum\limits_{i=1}^t f(\mathbf{s}^0_i)- F(\mathbf{x}^0)\right)\right|\\
 &~\leq~   \left|\dfrac{1}{t}\sum\limits_{i=1}^t f(\mathbf{s}^1_i)-F(\mathbf{x}^1)\right| + \left|\dfrac{1}{t}\sum\limits_{i=1}^t f(\mathbf{s}^0_i)- F(\mathbf{x}^0)\right|\\
  &~\leq~ 2\varepsilon_0|\max\limits_{\mathbf{s}\in \Delta_k^n}  f(\mathbf{s})| \\
  &~\leq~ 2\varepsilon_0nM,
\end{align*}
with probability at least $1 - 2e^{-t\epsilon^2/4}$. 
By setting $\varepsilon_0= \frac{\varepsilon}{2kn^3M}$ and $t  = \lceil\frac{16k^2n^6M^2\log\left( \frac{Mn^2+1}{\varepsilon\eta}\right)}{\varepsilon^2}\rceil$ we prove the lemma.}

\section{Proof of Lemma \ref{thm:properties}: Properties of the multilinear extension for $k$-submodular functions}
\label{app:property}

For ease of analysis, we introduce an equivalent definition of $k$-submodular functions.
Denote $(k+1)^V := \{(X_1,\ldots,X_k)\mid X_i \subseteq V,\forall i\in [k], X_i\cap X_j=\emptyset ,\forall i\neq j\}$ as the family of $k$ disjoint sets.
\begin{definition}[\bf{An equivalent definition of $k$-submodular functions}]
\label{def:k_sub_2}
A function $f\colon (k+1)^V\to \R$ is called \emph{$k$-submodular} if for any $S = (S_1, \dots , S_k)$ and $T = (T_1, \dots , T_k)$ in $(k+1)^V$,
\[
f(S)+f(T)\geq f(S\sqcap T)+f(S\sqcup T)\,,
\]
where 
\[
S\sqcap T = (S_1\cap T_1,\dots , S_k\cap T_k)
\]
and 
\[
S\sqcup T = ((S_1\cup T_1)\setminus \bigcup_{i\neq 1}(S_i\cup T_i), \dots, (S_k\cup T_k)\setminus \bigcup_{i\neq k}(S_i\cup T_i))\,.
\]
\end{definition}

\noindent
We also define the multilinear extension via this definition. 

\begin{definition}[\bf{Induced multilinear extension of Definition~\ref{def:k_sub_2}}]
\label{def:mul2}
Define $F\colon \Delta_k^n \to \R_{\geq 0}$ for a $k$-submodular function $f\colon (k+1)^V\to \R_{\geq 0}$ as
\begin{equation*}
F(\mbx) = \sum\limits_{S_1\uplus\dots\uplus S_k=S\subseteq V} f(S_1,\dots,S_k)  \Big(\prod\limits_{j\in [k]}\prod\limits_{i\in S_j}\mbx_{i,j}\Big) \prod\limits_{i\in V\setminus S}\Big(1-\sum\limits_{j=1}^k\mbx_{i,j}\Big)\,,
\end{equation*}
where $n=|V|$, ``$\uplus$'' denotes disjoint union.
\end{definition}

\noindent
This definition is identical to the Definition \ref{def:mul} by the equivalence of the definitions of $k$-submodularity shown in \cite{DBLP:journals/talg/WardZ16}.

We denote
\[\Delta_{e,i}f(X) = f(X_1,\dots,X_{i-1},X_i\cup \{e\},X_{i+1},\dots ,X_k) - f(X_1,\dots,X_k)\]
for $\mbx \in (k+1)^V, e \not\in \bigcup^{k}_{j=1}X_j$, and $i \in [k]$, which represents the marginal gain when adding $e$ to the $i$-th component of $X$. 
Then it is straightforward to notice that $k$-submodularity implies orthant submodularity
\[\Delta_{e,i}f(X)\geq \Delta_{e,i}f(Y),~\forall X,Y \in (k+1)^V \mbox{ with }X \le Y , e\not\in \bigcup^{k}_{l=1}Y_l, i \in [k]\,,\]
and pairwise monotonicity
\begin{equation}
\label{eqn:disc-pairwise-monotone}
\Delta_{e,i_1}f(X)+ \Delta_{e,i_2}f(X)\geq 0,~\forall X\in (k+1)^V \mbox{ with } e\not\in \bigcup^{k}_{l=1}X_l, i_1,i_2 \in [k], i_1 \neq i_2\,.
\end{equation}
Ward and \v Zivn\'{y} show that the converse is also true~\cite{DBLP:journals/talg/WardZ16}.
Functions that are submodular in every orthant and are pairwise monotone must be $k$-submodular.
Similar results are proved for bisubmodular functions earlier by Ando, Fujishige, and Naitoh \cite{ando1996characterization}.

Now we show the properties via Definition \ref{def:mul2} instead of Definition \ref{def:mul}.
\paragraph{Multilinearity.} Taking derivative of \eqref{eqn:multilinear} with respect to $\mbx_{i,j}$,
\begin{align*}
\frac{\partial F}{\partial \mbx_{i,j}}
=&\sum\limits_{i\in S_j} f(S_1,\dots,S_k)  \prod\limits_{t\in [k]}\prod\limits_{l\in S_t \atop l\neq i}\mbx_{l,t} \prod\limits_{l\in V\setminus S}\Big(1-\sum\limits_{t=1}^k\mbx_{l,t}\Big) \\
&- \sum\limits_{i\notin S} f(S_1,\dots,S_k)  \prod\limits_{t\in [k]}\prod\limits_{l\in S_t}\mbx_{l,t}  \prod\limits_{l\in V\setminus S \atop l\neq i}\Big(1-\sum\limits_{t=1}^k\mbx_{l,t}\Big)\,.
\end{align*}
As both the terms do not depend on $x_{i,j}$, the derivative is constant when other coordinates are fixed.
\paragraph{Element-wise non-positive Hessian.} Taking the second-order derivative of \eqref{eqn:multilinear} with respect to $\mbx_{i_1,j_1}$ and $\mbx_{i_2,j_2}$, 
\begin{align}
\label{eq:secder}
\begin{split}
\frac{\partial^2 F}{\partial \mbx_{i_1,j_1}\partial \mbx_{i_2,j_2}}
=&\sum\limits_{i_1\in S_{j_1} \atop i_2\in S_{j_2}} f(S_1,\dots,S_k)  \prod\limits_{t\in [k]}\prod\limits_{l\in S_t \atop l\neq i_1, i_2}\mbx_{l,t}  \prod\limits_{l\in V\setminus S}\Big(1-\sum\limits_{t=1}^k\mbx_{l,t}\Big)\\
&-\sum\limits_{i_1\notin S \atop i_2\in S_{j_2}} f(S_1,\dots,S_k)  \prod\limits_{t\in [k]}\prod\limits_{l\in S_t \atop l\neq i_2}\mbx_{l,t} \prod\limits_{l\in V\setminus S \atop l\neq i_1}\Big(1-\sum\limits_{t=1}^k\mbx_{l,t}\Big)\\
&-\sum\limits_{i_1\in S_{j_1} \atop i_2\notin S} f(S_1,\dots,S_k)  \prod\limits_{t\in [k]}\prod\limits_{l\in S_t \atop l\neq i_1}\mbx_{l,t} \prod\limits_{i\in V\setminus S \atop l\neq i_2}\Big(1-\sum\limits_{t=1}^k\mbx_{l,t}\Big)\\
&+\sum\limits_{i_1\notin S \atop i_2\notin S} f(S_1,\dots,S_k)  \prod\limits_{t\in [k]}\prod\limits_{l\in S_t}\mbx_{l,t} \prod\limits_{l\in V\setminus S \atop l\neq i_1,i_2}\Big(1-\sum\limits_{t=1}^k\mbx_{l,t}\Big)\,.
\end{split}
\end{align}
If $i_1\neq i_2$, for every subset tuple $S= S_1\uplus \dots \uplus S_k$ such that $i_1,i_2 \notin S$ which is in the fourth term, we can find a subset tuple $S^1$ such that 
\[S^1 = S_1\uplus\dots\uplus S_{j_1-1}\uplus (S_{j_1}\cup \{i_1\} )\uplus S_{j_1+1}\uplus \dots\uplus S_k\,,\]
in the third sum, and a subset tuple $S^2$ such that,
\[S^2 = S_1\uplus\dots\uplus S_{j_2-1}\uplus (S_{j_2}\cup \{i_2\} )\uplus S_{j_2+1}\uplus \dots\uplus S_k\,,\] 
in the second sum, and a subset tuple $S^0$ such that 
\begin{align*}
S^0 =~ &S_1 \uplus \dots \uplus S_{j_1-1} \uplus (S_{j_1}\cup \{i_1\} ) \uplus S_{j_1+1}\\
&~\uplus\dots\uplus  S_{j_2-1}\uplus (S_{j_2}\cup \{i_2\} )\uplus S_{j_2+1} \uplus \dots \uplus S_k\, ,
\end{align*}
assuming $j_1\leq j_2$ without loss of generality. Thus we have $S^1\sqcap S^2 = S$ and $S^1 \sqcup S^2 = S^0$. Due to submodularity, we have 
\[f(S^0)+f(S)-f(S^1)-f(S^2)\leq 0\,,\]
which implies that 
\[\frac{\partial^2 F}{\partial \mbx_{i_1,j_1}\partial \mbx_{i_2,j_2}}\leq 0\,.\]
{On the other hand, due to submodularity, we have 
\[|f(S^0)+f(S)-f(S^1)-f(S^2)|\leq |f(S^0)-f(S^1)|+|f(S^2)-f(S)|\leq 2M\,,\]
which implies that 
\[\frac{\partial^2 F}{\partial \mbx_{i_1,j_1}\partial \mbx_{i_2,j_2}}\geq -2M\,.\]
}
If $i_1= i_2= i$, by the multilinearity we have
\[
\frac{\partial^2 F}{\partial \mbx_{i,j_1}\partial \mbx_{i,j_2}}=0\,.
\]
\paragraph{Pairwise monotonicity.} Taking derivative of \eqref{eqn:multilinear},
\begin{align*}
\frac{\partial F}{\partial \mbx_{i,j_1}} + \frac{\partial F}{\partial \mbx_{i,j_2}}
=~&\sum\limits_{i\in S_{j_1}} f(S_1,\dots,S_k)  \prod\limits_{t\in [k]}\prod\limits_{l\in S_t \atop l\neq i}\mbx_{l,t} \prod\limits_{l\in V\setminus S}\Big(1-\sum\limits_{t=1}^k\mbx_{l,t}\Big) \\
&~- \sum\limits_{i\notin S} f(S_1,\dots,S_k)  \prod\limits_{t\in [k]}\prod\limits_{l\in S_t}\mbx_{l,t}  \prod\limits_{l\in V\setminus S \atop l\neq i}\Big(1-\sum\limits_{t=1}^k\mbx_{l,t}\Big)\\
&~\sum\limits_{i\in S_{j_2}} f(S_1,\dots,S_k)  \prod\limits_{t\in [k]}\prod\limits_{l\in S_t \atop l\neq i}\mbx_{l,t} \prod\limits_{l\in V\setminus S}\Big(1-\sum\limits_{t=1}^k\mbx_{l,t}\Big) \\
&~- \sum\limits_{i\notin S} f(S_1,\dots,S_k)  \prod\limits_{t\in [k]}\prod\limits_{l\in S_t}\mbx_{l,t}  \prod\limits_{l\in V\setminus S \atop l\neq i}\Big(1-\sum\limits_{t=1}^k\mbx_{l,t}\Big)\,.
\end{align*}
For every set tuple $S= S_1\uplus \dots \uplus S_k$ such that $i \notin S$, which is in the second term and the fourth term, we can find a set tuple $S^1$ such that 
\[
S^1 = S_1\uplus\dots\uplus S_{j_1-1}\uplus (S_{j_1}\cup \{i\} )\uplus S_{j_1+1}\uplus \dots\uplus S_k
\]
in the first term, and a set tuple $S^2$ such that 
\[
S^2 = S_1\uplus\dots\uplus S_{j_2-1}\uplus (S_{j_2}\cup \{i\} )\uplus S_{j_2+1}\uplus \dots\uplus S_k
\]
in the third term. By the (discrete) pairwise monotonicity \eqref{eqn:disc-pairwise-monotone} of $k$-submodular functions, we have 
\[
S^1-S+S^2-S = \Delta_{i,j_1}f(S)+\Delta_{i,j_2}f(S)\geq 0\,.
\]
Thus 
\[
\frac{\partial F}{\partial \mbx_{i,j_1}} + \frac{\partial F}{\partial \mbx_{i,j_2}}\geq 0\,.
\]
\paragraph{Approximate linearity.} Since $F$ is polynomial in $\mbx$, by {the Lagrangian form of} Taylor’s Theorem, $F(\mbx')$ at $F(\mbx)$ can be expanded as
\begin{align*}
~F(\mbx')-F(\mbx)  ~=  (\mbx'-\mbx)^T \nabla F(\mbx) + \dfrac{1}{2} (\mbx'-\mbx)^TH(\xi)(\mbx'-\mbx),
\end{align*}
{where $H(\cdot)$ is the Hessian matrix, and $\xi$ is a point that lies on the line segment connecting points $\mbx $ and $\mbx'$,}
%
Now we consider an element $\frac{\partial^2 F{(\xi)}}{\partial \mbx_{i_1,j_1}\partial \mbx_{i_2,j_2}}$ in $H(\xi)$. 
%
{By the property of Element-wise non-positive Hessian, we have 
}
{
\[
\left|\frac{\partial^2 F}{\partial \mbx_{i_1,j_1}\partial \mbx_{i_2,j_2}}\right| \leq 2M.   
\]
Therefore, if $\mbx' -\mbx \in \delta\Delta_k^n$, i.e., for all $i \in [n]$, $\sum\limits_{j\in [k]}\mbx'_{i,j} -\mbx_{i,j}\leq \delta$, we have 
\begin{align*}
&\left|\dfrac{1}{2} (\mbx'-\mbx)^TH(\xi)(\mbx'-\mbx) \right|\\
~\leq ~&  \sum\limits_{i_1\in [n],j_1\in [k]}\sum\limits_{i_2\in [n],j_2\in [k]}
\left|\frac{\partial^2 F}{\partial \mbx_{i_1,j_1}\partial \mbx_{i_2,j_2}}\right|
|\mbx'_{i_1,j_1}-\mbx_{i_1,j_1}||\mbx'_{i_2,j_2}-\mbx_{i_2,j_2}|\\
~\leq ~& \dfrac{1}{2} \sum\limits_{i_1\in [n],j_1\in [k]}\sum\limits_{i_2\in [n],j_2\in [k]}2M
|\mbx'_{i_1,j_1}-\mbx_{i_1,j_1}||\mbx'_{i_2,j_2}-\mbx_{i_2,j_2}|\\
~= ~&M \sum\limits_{i_1\in [n]}\sum\limits_{i_2\in [n]}\left(\sum\limits_{j_1\in [k]}
|\mbx'_{i_1,j_1}-\mbx_{i_1,j_1}|\right)\left(\sum\limits_{j_2\in [k]}|\mbx'_{i_2,j_2}-\mbx_{i_2,j_2}|\right)\\
~\leq ~&  M\sum\limits_{i_1\in [n]}\sum\limits_{i_2\in [n]}
\delta^2\\
~=~& n^2\delta^2M.
\end{align*}
Therefore,
\begin{align*}
 F(\mbx')-F(\mbx)\geq~&\sum\limits_{i\in[n],j\in [k] } (\mbx'_{i,j} -\mbx_{i,j} )\cdot \partial_{i,j} F(\mbx)-{ n^2\delta^2M}
 .  
\end{align*}
}
\paragraph{Preservation of monotonicity.}
As 
\begin{align*}
\frac{\partial F}{\partial \mbx_{i,j}}
=&\sum\limits_{i\in S_j} f(S_1,\dots,S_k)  \prod\limits_{t\in [k]}\prod\limits_{l\in S_t \atop l\neq i}\mbx_{l,t} \prod\limits_{l\in V\setminus S}\Big(1-\sum\limits_{t=1}^k\mbx_{l,t}\Big) \\
&- \sum\limits_{i\notin S} f(S_1,\dots,S_k)  \prod\limits_{t\in [k]}\prod\limits_{l\in S_t}\mbx_{l,t}  \prod\limits_{l\in V\setminus S \atop l\neq i}\Big(1-\sum\limits_{t=1}^k\mbx_{l,t}\Big)\,,
\end{align*}
for every set tuple $S= S_1\uplus \dots \uplus S_k$ such that $i \notin S$, which is in the second term, we can find $S'$ such that 
\[
S' = S_1\uplus\dots\uplus S_{j-1}\uplus (S_j\cup \{i\} )\uplus S_{j+1}\uplus \dots\uplus S_k
\]
in the first term. When $f$ is monotone (assume that $f$ is increasing without loss of generality), we have $f(S')\geq f(S)$. Thus $\frac{\partial F}{\partial \mbx_{i,j}}\geq 0$, for all $i\in [n]$ and $j\in [k]$, which indicates that $F$ is also monotone as desired.

\section{Proof of Lemma \ref{lmm:rounding}: A novel rounding scheme}
\label{sec:rounding}

In this section, we prove Lemma \ref{lmm:rounding} by describing and analyzing a rounding algorithm called $\mathtt{KSUBROUND}$ (Algorithm \ref{alg:ksubround}).

\paragraph{Useful notations and facts for Lemma \ref{lmm:rounding}.}
We first recall the round procedure of submodular functions~\citep{DBLP:journals/siamcomp/CalinescuCPV11,DBLP:journals/siamcomp/ChekuriVZ14}.
\begin{lemma}[\citep{DBLP:journals/siamcomp/CalinescuCPV11,DBLP:journals/siamcomp/ChekuriVZ14}]
\label{lmm:subround}
Assume $\mathcal{P}\subseteq [0,1]^n$ is a down-closed polytope and $G:[0,1]^n\to \mathbb{R}_{\geq 0}$ is a multilinear extension of some submodular function $g$, then
there exists an algorithm, that takes a vector $\mathbf{y}\in \mathcal{P}$ and the function $G$ as input and return a set $S\in 2^n$ obeying $\mathrm{1}_{S}\in \mathcal{P}$ and  
\begin{itemize}
\item $\mathbb{E}\left[G(\mathrm{1}_{S})\right]\geq G(\mathbf{y})$ {without any query to $g$} , when $\mathcal{P}$ is single matroid constraint;
    \item $\mathbb{E}\left[G(\mathrm{1}_{S})\right]\geq (1-\varepsilon)G(\mathbf{y})$ for any fixed $\varepsilon > 0$ {with query complexity $O(n^{poly(1/\varepsilon}))$ to $g$} , when $\mathcal{P}$ is $l=O(1)$ knapsack constraints;
    \item $\mathbb{E}\left[G(\mathrm{1}_{S})\right]\geq \left(\frac{0.6}{b}-\varepsilon\right)G(\mathbf{y})$ {for any fixed $\varepsilon > 0$ with query complexity $O(n^{poly(1/\varepsilon}))$ to $g$} , when $\mathcal{P}$ is intersection of $b$ matroid constraints and $l=O(1)$  knapsack constraints.
\end{itemize}
with polynomial times queries to $G$. We refer this algorithm as $\mathtt{SUBROUND}(\mathbf{y},G)$.
\end{lemma}

%
Our approach uses the rounding procedures $\mathtt{SUBROUND}$, which are applied after reducing the multilinear extension of the $k$-submodular function to the multilinear extension of a submodular function with an index vector $\mathbf{I}\in \{1,\ldots,k\}^n$.

\sloppy
\begin{definition}[\bf{Reduced multilinear extension}]
Given a multilinear extension of $k$-submodular function, $F:\Delta_k^n\to \mathbb{R}_{\geq 0}$ for any index vector $\mathbf{I}\in \{1,\ldots,k\}^n$, we define a reduced function $F_{\mathbf{I}}:[0,1]^n\to \mathbb{R}_{\geq 0}$ as 
\[F_{\mathbf{I}}({\mathbf{x}}) = F({\mathbf{x}}^\mathbf{I}),\]
where ${\mathbf{x}}^\mathbf{I}\in \Delta^n_k$ is defined as 
\[{\mathbf{x}}^\mathbf{I}_{i,j}= \left\{\begin{array}{cc}
\mathbf{x}_i    &  j =\mathbf{I}_i, \\
 0    & \mbox{otherwise}.
\end{array}\right.\]
\end{definition}
Intuitively, we define a reduced function by constraining ${\mathbf{x}}^\mathbf{I}$ to only take non-zero values at the coordinates specified by an index vector $\mathbf{I}\in \{1,\ldots, k\}^n$.
%
%
Such reduced functions enjoy the submodularity shown in Claim \ref{clm:reduce}.
\begin{claim}
\label{clm:reduce}
If $F$ is a multilinear extension of the $k$-submodular function $f$, the reduced function $F_\mathbf{I}$ is a multilinear extension of the submodular function $f_\mathbf{I}: 2^n\to \mathbb{R}_{\geq 0}$ defined as 
\[f_{\mathbf{I}}(S) = f({S}^\mathbf{I}),\]
where ${S}^\mathbf{I}\in \{0,\ldots,k\}^n$ is defined as 
\[{S}^\mathbf{I}_{i}= \left\{\begin{array}{cc}
\mathbf{I}_i   &  i\in S, \\
 0    & i\not\in S.
\end{array}\right.\]
\end{claim}

\begin{proof}
We first illustrate the function-extension correspondence in the following figure.
\[\begin{tikzcd}
f \arrow[rr, "\text{Multilinear}","\text{Extension}"'] \arrow[dd,"\mathbf{I}"] &  & F\arrow[dd,"\mathbf{I}"] \\
                                          &  &                         \\
f_\mathbf{I} \arrow[rr, "\spadesuit~\text{Multilinear}","\text{Extension}"']               &   & F_\mathbf{I}            
\end{tikzcd}  \]
We also illustrate the domain correspondence in the below figure.
\[\begin{tikzcd}
\{0,\ldots,k\}^n \arrow[rr, "\text{Extension}"'] \arrow[dd,"\mathbf{I}"] &  & \Delta_k^n \arrow[dd,"\mathbf{I}"]\\
                                          &  &                         \\
2^n\arrow[rr, "\text{Extension}"']               &   & {[0,1]}^n           
\end{tikzcd}  \]

In the following, we complete the proof by showing the submodularity of  $f_\mathbf{I}$ and prove the multilinear extension relationship between $f_\mathbf{I}$ and $F_\mathbf{I}$ (marked as $\spadesuit$ in the first figure).

We obtain the submodularity of $f_\mathbf{I}$ by the inequality that 
\[f_\mathbf{I}(S)+f_\mathbf{I}(T) = f({S}^\mathbf{I}) + f({T}^\mathbf{I})\geq f(\min_0({S}^\mathbf{I},{T}^\mathbf{I}))+f(\max_0({S}^\mathbf{I},{T}^\mathbf{I})) =f_\mathbf{I}(S\cap T)+f_\mathbf{I}(S\cup T). \]
By the definition of the reduced function, we have 
\begin{align*}
 F_\mathbf{I}(\mbx) =   ~&  F({\mathbf{x}}^\mathbf{I}) \\
 =~&\sum\limits_{\mathbf{s}\in \{0,\ldots,k\}^{n}} f(\mathbf{s})  \prod\limits_{i \in [n]: \mathbf{s}_i \neq 0}\mbx^\mathbf{I}_{i,\mathbf{s}_i}\prod\limits_{i\in [n]:\mathbf{s}_i = 0}\Big(1-\sum\limits_{j=1}^k\mbx^\mathbf{I}_{i,j}\Big)\\
  =~&\sum\limits_{S\in 2^n} f(S^\mathbf{I})  \prod\limits_{i \in [n]: S^\mathbf{I}_i \neq 0}\mbx^\mathbf{I}_{i,S^\mathbf{I}_i} \prod\limits_{i\in [n]:S^\mathbf{I}_i = 0}\Big(1-\sum\limits_{j=1}^k\mbx^\mathbf{I}_{i,j}\Big)\\
   =~&\sum\limits_{S\in 2^n} f(S^\mathbf{I})  \prod\limits_{i\in [n]: S^\mathbf{I}_i  \neq 0}\mbx_i \prod\limits_{i\in [n]:S^\mathbf{I}_i = 0}\Big(1-\mbx_i\Big)\\
    =~&\sum\limits_{S\in 2^n} f_\mathbf{I}(S)  \prod\limits_{ i\in S}\mbx_i\prod\limits_{i\not\in S}\Big(1-\mbx_i\Big).
\end{align*}
Thus, $F_\mathbf{I}$ is the multilinear extension of $f_\mathbf{I}$.

The correspondence between the reduced function and the submodular function can also be understood through a probabilistic view. 
Specifically, we consider a random vector $\Tilde{\mathbf{s}}\in \{0,\ldots,k\}^n$, where each entry $\Tilde{\mathbf{s}}_i\neq 0$ is drawn independently with probability $\sum\limits_{j=1}^{n} \mbx^\mathbf{I}_{i,j}$ for each $i\in [n]$, and we have $\sum\limits_{j=1}^{n} \mbx^\mathbf{I}_{i,j} = \mbx^\mathbf{I}_{i,\mathbf{I}_i} =\mbx_i $.
If $\Tilde{\mathbf{s}}_i\neq 0$  in this process, $\mathbf{s}_i$ is assigned with value $\mathbf{I}_i$.
We can observe that this probability-based definition of reduced function is equivalent to the probability-based definition of the multilinear extension.

\end{proof}

\paragraph{Algorithm for Lemma \ref{lmm:rounding}.}
Now we are ready to introduce our rounding algorithm $\mathtt{KSUBROUND}$ (Algorithm \ref{alg:ksubround}) which consists of  three phases: rounding from $\Delta_k^n$ to $[0,1]^n$ (Lines 1-7), rounding from $[0,1]^n$ to $2^n$ (Line 8) and recovering from $2^n$ to $(k+1)^n$ (Lines 9-12).
In each phase, we preserve the feasibility and control the loss.

In the first phase (Line 1-7), for any $i$, we merge all non-zero values and assign value $j$ to the $i$-th coordinate of the index vector $\mathbf{I}$ with a categorical probability of the proportion.
This merging process does not conflict with the support constraint because the sum $\sum_{j=1}^k\mbx_{i,j}$ remains constant. 
Furthermore, at each iteration $i$, by the definition of multilinear extension $F$, the function value is exactly the linear combination of function value at every vertex of the affined corner of cube, i.e., 
\footnote{We can also conclude the linearity by the zero value of the Hessian element at the same element $i$'s block, i.e., $\frac{\partial^2 F}{\partial x_{i,j_1}\partial x_{i,j_2}}=0$ for any $j_1,j_2\in [k]$.}
\begin{align*}
 \mathbb{E}_{\mathbf{I}}\left[F_\mathbf{I}(\mathbf{y})\right] ~=~& \mathbb{E}_{\mathbf{I}}\left[F(\mby^\mathbf{I})\right]\\
 ~=~&\mathbb{E}_{\mathbf{I}}\left[ \sum\limits_{\mathbf{s}\in \{0,\ldots,k\}^{n}} f(\mathbf{s})  \prod\limits_{i \in [n]: \mathbf{s}_i \neq 0}\mby^\mathbf{I}_{i,\mathbf{s}_i}\prod\limits_{i\in [n]:\mathbf{s}_i = 0}\Big(1-\sum\limits_{j=1}^k\mby^\mathbf{I}_{i,j}\Big)\right]\\
 ~=~& \sum\limits_{\mathbf{s}\in \{0,\ldots,k\}^{n}} f(\mathbf{s})  \prod\limits_{i \in [n]: \mathbf{s}_i \neq 0}\mbx_{i,\mathbf{s}_i}\prod\limits_{i\in [n]:\mathbf{s}_i = 0}\Big(1-\sum\limits_{j=1}^k\mbx_{i,j}\Big)\\
 ~=~& F(\mbx).
\end{align*}

In the second phase (Line 14), we apply the rounding procedure $\mathtt{SUBROUND}$ which takes the fractional solution $\mathbf{y} \in \mathcal{P}\subseteq [0,1]^n$ and the reduced function $F_{\mathbf{I}}$ as input and returns 
an integer solution $S\in 2^n$.
The loss of this rounding procedure is bounded by Lemma \ref{lmm:subround}.

In the final phase (Lines 16-18), we recover the solution $\mathbf{s}\in \Delta_k^n$ by setting $\mathbf{s}_i=\mathbf{I}_i$ if $i\in S$.
This recovery step incurs no loss since the recovery procedure and reducing procedure correspond to the same index vector $\mathbf{I}$.

\begin{algorithm}[ht]
 \caption{$\mathtt{KSUBROUND}$}
  \label{alg:ksubround}
\DontPrintSemicolon
\SetNoFillComment
  \SetKwInOut{Input}{Input}
  \Input{A fractional solution $\mathbf{x}\sim \mathcal{P}\subseteq \Delta_k^n$, $\mathcal{O}_{F,\nabla F}$, membership oracle of $\mathcal{P}$.}
{\bf{Initialize}} $\mathbf{y} \leftarrow [0,\ldots,0]^\top\in [0,1]^n $ and $\mathbf{I}\leftarrow [0,\ldots,0]^\top\in \{0,\ldots,k\}^n$.\;
 \For{$i \in [n]$}{
    $\mathbf{y}_i \leftarrow \sum\limits_{j=1}^k\mathbf{x}_{i,j} $.\;
    \If{$\mathbf{y}_i\neq 0$}{
    \textbf{With Categorical Probability $p = {{\mathbf{x}}_{i,j}}/{\sum\limits_{j=1}^k{\mathbf{x}}_{i,j}}$}: $\mathbf{I}_i \leftarrow j$.\;
    }
    \Else{
    $\mathbf{I}_i \leftarrow 0$.
    }
    }
$S\leftarrow\mathtt{SUBROUND}(\mathbf{y},F_{\mathbf{I}})$.\;
Initialize $\mathbf{s}\leftarrow [0,\ldots,0]^\top\in \{0,\ldots,k\}^n$.\;
\For{$i\in [n]$}{
\If{$i\in S$}{
$\mathbf{s}_i \leftarrow \mathbf{I}_i$.
}
}
\textbf{Return:} $\mathbf{s}$.
\end{algorithm}

\begin{proof}[Proof of Lemma~\ref{lmm:rounding}]
We first analyze the feasibility and then prove the approximation performance.

\textbf{Feasibility}
In Lines 1-7, since $\mbx\in \mathcal{P}^c = \left\{\mbx\in \Delta_k^n:\left(\sum\limits_{j=1}^k\mbx_{1,j},\ldots,\sum\limits_{j=1}^k\mbx_{n,j}\right)^\top\in \mathcal{P} \right\}$,
and the sum $\sum\limits_{j=1}^k\mbx_{i,j}$ will remains constant during the moving for any $i$, we have $\widehat{\mbx}\in \mathcal{P}^c$ and $\mathbf{y}\in \mathcal{P}$.
In Line 8, by Lemma \ref{lmm:subround}, we have $1_S\in \mathcal{P}$. In Line 9-12, by the definition of $\mathcal{P}^c$, we have $\mathbf{s}\in  \mathcal{P}^c$, i.e., $\mathbf{s}\sim \mathcal{P}$.

\textbf{Approximation ratio of Algorithm \ref{alg:ksubround}.}
In Line 1-7, by the definition of $F$ and $F^\mathbf{I}$, we conclude that 
\begin{equation}
\label{eq:round1}
\mathbb{E}_{\mathbf{I}}\left[F^\mathbf{I}(\mathbf{y})\right] =  F(\mathbf{x}).
\end{equation}
In Lines 9-12, by the definition of $F_{\mathbf{I}}$, we have 
\begin{equation}
\label{eq:round3}
F(1_{\mathbf{s}}) = F_{\mathbf{I}}(1_{S}).
\end{equation}
Combine Eq. \eqref{eq:round1} and \ref{eq:round3} and Lemma \ref{lmm:subround}, we complete the proof.
\end{proof}


\section{Hardness for the intersection of $O(1)$ knapsacks and $b$ matroids}
\label{sec:hardness}


\begin{theorem}[\bf{Hardness for the intersection of $b$ matroids}]
\label{thm:hardness}
There exist instances of $k$-submodular maximization with support constraints, $\max \{f (\mathbf{s}),\mathbf{s}\in \mathcal{P}\}$, where $\mathcal{P}$ is intersection of $b$ matroids, any algorithm with better than $O(\log b/b+\varepsilon)$ approximation ratio
for this problem would require exponentially many value queries for any $\varepsilon > 0$, unless $P=NP$.
\end{theorem}

\begin{proof}
Consider the monotone $k$-submodular function $g:\{0,1,2\}^n\to \mathbb{R}_{\geq 0}$, defined as 
\[g(\mathbf{s}) = f(S_1)+ \varepsilon f(S_2),\]
where $f:2^n\to \mathbb{R}_{\geq 0}$ is a monotone submodular function, $\varepsilon\in\mathbb{R}^+ $, $S_1 = \{i:\mathbf{s}_i = 1\}\in 2^n$ and $S_2 = \{i:\mathbf{s}_i = 2\}\in 2^n$.
We set $\varepsilon$ to be sufficiently small such that 
\[\min\limits_{i}f(N)-f(N\setminus\{i\})\geq\varepsilon\left( \max\limits_{i}f(i)-f(\varepsilon)\right).\]
This ensures that the optimal solution $\mathbf{o}\in \{0,1,2\}^n$ satisfies that $\mathbf{o}_i\in \{0,1\}$ for all $i\in [n]$. Otherwise, we can improve the function value by changing the value $\mathbf{o}_i$ from $2$ to $1$ without conflict with the support constraint. 
In other words, maximizing the $k$-submodular function $g$ subject to any support constraint is equivalent to maximizing the submodular function $f$ with the same constraint.

It is well-known that unless $P=NP$, there is no approximation algorithm better than $O(\log b/b)$ for $b$-dimensional matching (see \citep{DBLP:journals/cc/HazanSS06}). Hence, there is no better approximation algorithm for submodular maximization subject to the intersection of $b$ matroids constraint \citep{DBLP:journals/mor/LeeSV10}. Therefore, the hardness result of $O(\log b/b)$ also holds for constrained $k$-submodular maximization.   
\end{proof}

\end{document}